\documentclass[english,12pt]{article}

\usepackage[T1]{fontenc}
\usepackage[latin9]{inputenc}
\usepackage{babel}
\usepackage{hyperref}
\usepackage{graphicx}

\usepackage[usenames]{color}
\definecolor{deanPURPLE}{rgb}{.3,0,.5}

\newcommand{\OR}{OR-star} % make sure we can search for the name if we
\newcommand{\impact}{{\color{red}b}}
\newcommand{\m}{\textcolor{blue}{m}}
\newcommand{\flux}{\textcolor{red}{\tilde{m}}}
\newcommand{\ALTdensity}{\textcolor{red}{\hat{m}}}

\usepackage{amsmath}
\usepackage{amsthm}
\usepackage{amssymb}

\DeclareMathOperator{\tr}{tr}

\newcommand{\Einf}{E^{\infty}}% Use only in math mode
\newcommand{\Eloc}{E^{\text{loc}}}% Use only in math mode

\providecommand{\abs}[1]{\lvert #1 \rvert}

\newtheorem{lemma}{Lemma}
\newtheorem{theorem}{Theorem}

\begin{document}

\title{Orbiting Radiation Stars}
\author{Dean P. Foster\footnote{Current affiliation SCOT, Amazon.com,
NYC.  This work was done while he was visiting MSR, NYC from the University of
Pennsylvania.} and John Langford\footnote{MSR} and Gabe Perez-Giz\footnote{Current affiliation: Policy Fellow
  for the American Association for the Advancement of Science. This
  work was done while at the NYU Center for Cosmology and Particle
  Physics (NYU CCPP)}}

\maketitle
\begin{abstract}
  We study a spherically symmetric solution to the Einstein equations
  in which the source, which we call an \OR{} (orbiting radiation
  star), is a compact object consisting of freely falling null
  particles. The solution avoids quantum scale regimes and hence
  neither relies upon nor ignores the interaction of quantum mechanics
  and gravitation.  The \OR{} spacetime exhibits a deep gravitational
  well yet remains singularity free.  In fact, it is geometrically
  flat in the vicinity of the origin, with the flat region being of
  any desirable scale.  The solution is observationally distinct from
  a black hole because a photon from infinity aimed at an \OR{}
  escapes to infinity with a time delay.
\end{abstract}

\section{The problem}
\label{sec:intro}

Can photons form a star?  Or, more generally, can a collection of null
particles that interact only via their mutual gravitation form a bound
object?

Consider first a simpler setting: a swarm of Newtonian particles
occupying spherically and temporally symmetric orbits.  The classical
gravity felt by a particle at radius $r$ is due only to whatever mass
is inside radius $r$. Therefore:
\begin{enumerate}
\item If the swarm has an inner edge $r_{\text{inner}}$, then a
  particle at $r_{\text{inner}}$ must be moving tangent to that edge
  since the particle experiences no gravity and continues to move
  (instantaneously) along its straight tangent line to a larger
  radius.
\item As the particle's radial position increases, the amount of mass
  enclosed at smaller $r$ values also increases, causing the
  particle's path to curve.  If the mass inside the particle's radius
  becomes sufficient, the trajectory eventually curves enough to
  deflect the particle inwards at some maximum radius
  $r_{\text{outer}}$, repeating this cycle.
\end{enumerate}

\begin{figure}[b!]
\includegraphics{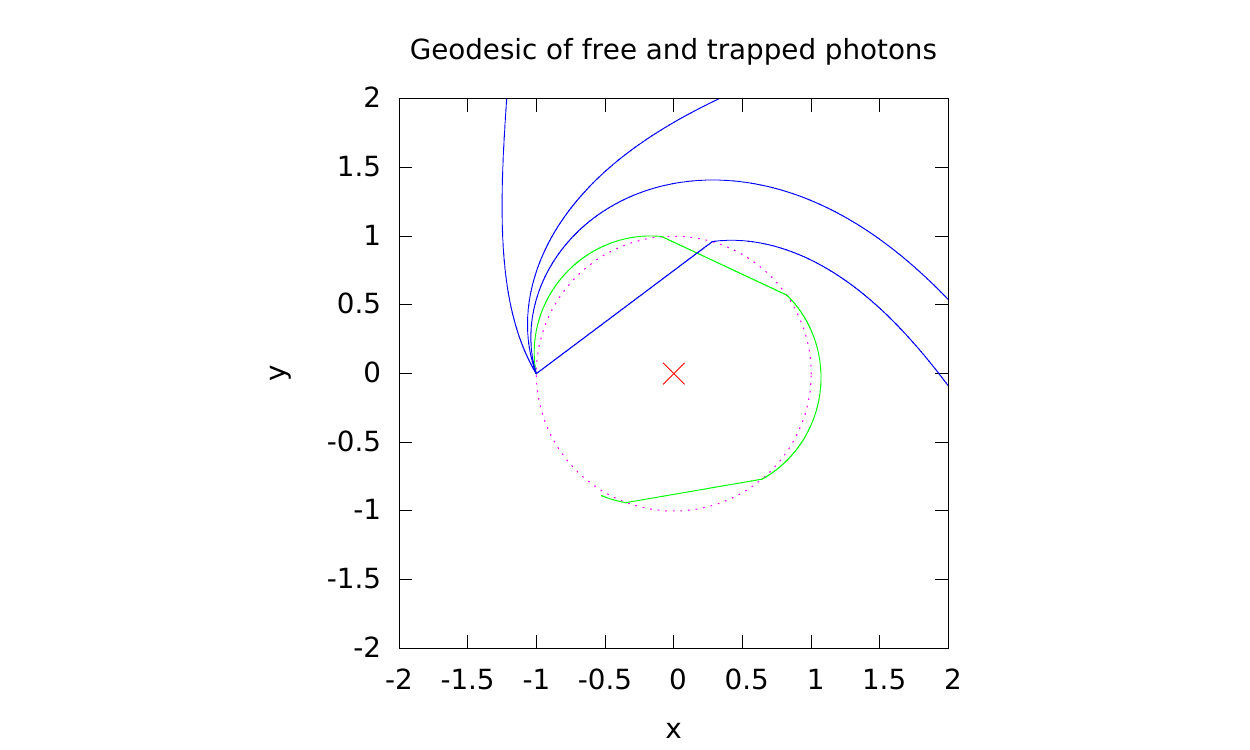}
\caption{A plot showing photons (null test particles) in the spacetime
  of a high-flux \OR{} with a center at the origin.  The dotted circle
  represents both the (almost degenerate) inner and outer radius of a
  high flux \OR{}.  Photons start at the edge of the flat inner radius
  with various orientations.  The blue lines show scattered photons
  that escape to infinity.  The green lines show trapped photon
  trajectories that repeatedly cross the the \OR{} surface in both
  directions.}
\label{fig:trapped}
\end{figure}

This intuition carries over to a relativistic version with a swarm of
null particles, each of which follows a bound null geodesic of the
spacetime generated by the swarm itself. We call the resulting object
an ``Orbiting Radiation star'' (or \OR{}). In this paper, we
demonstrate the existence of \OR{} solutions to the Einstein equations
and discuss their properties.  Figure~\ref{fig:trapped} probes an
\OR{} solution with test-particle photons to illustrate the geometry
of the solution.

\OR{}s differ quantitatively from but share several qualitative
features with the Newtonian scenario above.  The relativistic null
swarm has both an inner radius $r_{\text{inner}}$ and an outer radius
$r_{\text{outer}}$.  The inner radius can be of arbitrary real
dimension, with a flat spacetime geometry when $r < r_{\text{inner}}$.
At the outer radius, which can be significantly larger than the inner
radius, the solution matches onto an exterior Schwarzschild
geometry. In the \OR{} region
$r_{\text{inner}} \leq r \leq r_{\text{outer}}$, the null particles
follow planar eccentric bound null geodesics arranged in a spherically
symmetric configuration. As with most eccentric bound relativistic
orbits, these orbits are not simple ellipses, although they do have
well-defined pericenters and apocenters at $r_{\text{inner}}$ and
$r_{\text{outer}}$, respectively.

To preview some results, the minimum outer radius of an \OR\ is 9/8 of
its Schwarzschild radius.  And, unlike with Schwarzschild black holes,
the interior geometry of an \OR\ can be probed by observing the
deflection pattern and propagation delay of test particle photons sent
toward the \OR.

\subsection{Contrast with other solutions}

To elucidate some features of \OR{}s, it is helpful to contrast them
with two other families of solutions to the Einstein equations that
exhibit compact objects with deep gravitational wells.

\subsubsection{Black holes and similar}
The conventional Schwarzschild black hole solution to the Einstein
equations has a singularity at the origin and an event horizon at the
Schwarzschild radius.  Both of these elements ignore the potential for
quantum phenomena to modify the solution in the neighborhood of these
spacetime regions.

Gravastars~\cite{Grava2001, Grava2004} are black hole alternatives
that rely on Planck length scale effects to resist gravitational
collapse. Boson stars~\cite{BS86,BS92} rely on low-mass bosons to form
compact objects.  For particularly large objects, the wavelength of
the boson constituents must be particularly large.  Related to boson
stars are Geons~\cite{Geons}.  Spherically symmetric Geons are known
to be unstable, since a lower energy configuration exists with two
concentric counter-rotating rings ~\cite{Tolman}.  The timescale of
this instability, however, is unclear and almost certainly depends
inversely on the size of the object.  All of these solutions can be of
arbitrary scale, similar to an \OR{}, but either rely on quantum
effects or ignore them in regimes where quantum effects plausibly
matter.

In contrast, \OR{}s do not rely on quantum effects or ignore them in
regimes where they matter.  Its constituents can be any null particle
that interacts on large scales only via gravitational effects.  These
particles can have any energy distribution so long as the energy of
any single constituent particle is a negligible fraction of the
mass-energy of the entire ensemble.  The \OR{} has a well-defined
continuum limit as an anisotropic but otherwise ideal fluid, as
discussed in Section \ref{sec:anisotropicfluid}.

\subsubsection{Other null particle solutions}

Many null particle solutions to the Einstein
equations~\cite{crossstreaming,Gergeley,PhotonStars,KC2010,Vaidya}
have been studied.  Amongst all these solutions, the scenario of
colliding radially ingoing and outgoing null particle
streams~\cite{Gergeley} is a limiting case of the \OR{} solution with
some substantial technical similarities discussed in
Section~\ref{sec:neg-mass}.

This solution, however, requires a negative mass at the
origin. Although \OR{} solutions do also admit a negative central mass
and can use that feature to enhance their similarity to Schwarzschild
black holes, \OR{} solutions do not \emph{require} exotic negative
mass anywhere and in fact can join onto a flat spacetime interior
solution.  See section~\ref{sec:neg-mass} for details.

\subsection{Outline}

The rest of this paper is organized as follows. Section
\ref{sec:einsteintensor} lays out preliminaries like coordinate
choices and the form of the Einstein tensor for our problem. In
Section \ref{sec:stress-energy}, we derive the stress-energy tensor
for an \OR{}. In Section~\ref{sec:einsteineqns}, we reduce the
Einstein equations to a set of coupled ordinary differential equations
for the \OR{} geometry and for the geodesics of the null particles
that comprise it, deferring several computational details to
appendices.

In section~\ref{sec:simulation}, we discuss various approaches for
numerical integration of those equations.  We have tried them all and
have verified that their results agree within numerical error (we have
made all our integration codes available).

In Section~\ref{sec:results}, we discuss the structure of the solution
from several viewpoints. Finally, in Section~\ref{sec:variations}, we
discuss variations on the solution family which can be easily handled
and their implications.

\section{The Einstein tensor}
\label{sec:einsteintensor}

The solution we seek is spherically symmetric and static, so we use radial
Schwarzschild coordinates with the following notation for the metric:
\begin{eqnarray}
\label{eq:metric}
ds^2 = - A(r) dt^2 + B(r) dr^2 + r^2(d\theta^2 + \sin^2(\theta)
  d\phi^2) \, .
\end{eqnarray}
The metric functions $A(r)$ and $B(r)$ should match onto their
Schwarzschild values at the outer edge $r = r_{\text{outer}}$ of the
\OR{}.

Because an \OR{} is made up of null particles, its total stress-energy
tensor will be traceless (see equations
(\ref{eq:Tmunuanisotropicinrestframe})--(\ref{eq:anisotropic_fluid_eos})). The
Einstein tensor must therefore also be traceless, which implies that
the Ricci scalar vanishes identically.  As a result, just as in vacuum
spacetimes, the Einstein tensor in an \OR{} spacetime is simply the
Ricci tensor, which in these coordinates\footnote{See, for example,
  \cite{Zee}. Page 363 of this textbook derives these equations using
  the same notation we use here.} has nonvanishing components
\begin{subequations}
  \label{eq:Riccitensor}
  \begin{eqnarray}
    R_{tt} &=& \frac{A''}{2B} + \frac{A'}{rB} - \frac{A'}{4B}\left(\frac{A'}{A} + \frac{B'}{B}\right) \\
    R_{rr} &=& -\frac{A''}{2A} + \frac{B'}{rB} + \frac{A'}{4A}\left(\frac{A'}{A} + \frac{B'}{B}\right) \\
    R_{\theta\theta} &=& 1 - \frac{1}{B}  - \frac{r}{2B}\left(\frac{A'}{A}
                         - \frac{B'}{B}\right) \\
    R_{\phi\phi} &=& \sin^2(\theta) R_{\theta\theta}
  \end{eqnarray}
\end{subequations}

\section{The stress-energy tensor}
\label{sec:stress-energy}

The particles comprising an \OR{} follow null geodesics of the
spacetime generated by those same null particles.  A description of
those null geodesics allows us to deduce the form of the stress-energy
tensor of the entire OR star.  When all those geodesics have the same
magnitude impact parameter $\abs{b}$ and are distributed in a
spherically symmetric way, the resulting \OR{} can be interpreted as
an \emph{anisotropic} fluid with a distinct radial and tangential
pressure at every point.

% An \OR\ is spherically symmetric but with anisotropic pressure due to
% null particles imposing varying pressure in different directions
% rather than uniform pressure in all directions.  Handling this form of
% anisotropy requires only a minor modification of standard spherically
% symmetric isotropic proessure solutions.

\subsection{Null geodesics of \OR{} particles}
\label{sec:geodesics}

We derive the equations of motion along the fiducial null geodesic of
a single \OR{} particle.  The trajectory and 4-velocity of any other
particle can be derived from this fiducial geodesic by some
combination of rotation within the orbital plane, rotation of the
orbital plane, and a translation of orbital phase.

Spherical symmetry lets us choose the equatorial plane
$\theta \equiv \pi/2$ to be the orbital plane of the fiducial geodesic
without loss of generality.  $\partial_{\phi}$ is then a Killing
vector, so a particle with 4-momentum $p^{\mu}$ has a conserved
angular momentum
\begin{equation}
  \label{eq:angmomdefn}
  L \equiv g_{\mu\nu} \partial_{\phi}^{\mu} p^{\nu}
\end{equation}
that we allow to take any sign.  Because it is static, an \OR{}
spacetime also has a timelike Killing vector $\partial_{t}$ that
yields a conserved ``energy at infinity''
\begin{equation}
  \label{eq:Einfdefn}
  \Einf \equiv -g_{\mu\nu} \partial_{t}^{\mu} p^{\nu} \,.
\end{equation}

We define the impact parameter of each particle to be
\begin{equation}
  \label{eq:bdefn}
  \impact \equiv \frac{L}{\Einf}
\end{equation}
and allow it to take any sign (matching the sign of $L$ and
$\dot{\phi}$).  Because two null particles with the same $\impact$ but
different energies still follow the same geodesic, it is useful
to define an affine parameter $\lambda$ along null geodesics
\begin{equation}
  \label{eq:affineparamdefn}
  \frac{dx^{\mu}}{d\lambda} \equiv u^{\mu} \equiv
  \dot{x}^{\mu} \equiv
  \frac{1}{\Einf} p^{\mu}
\end{equation}
that absorbs this energy factor so that it need not appear explicitly
in the equations of motion.

Equation (\ref{eq:Einfdefn}) then yields the equation for $\dot{t}$,
\begin{equation}
  \label{eq:tdoteom}
  \Einf = - (-A(r)) \Einf \dot{t}
  \longrightarrow
  \dot{t} = \frac{1}{A(r)}
  \, ,
\end{equation}
while (\ref{eq:angmomdefn}) yields $\dot{\phi}$,
\begin{equation}
  \label{eq:phidoteom}
  L = r^{2} \Einf \dot{\phi}
  \longrightarrow
  \dot{\phi} = \frac{b}{r^{2}}
  \, .
\end{equation}
For the fiducial geodesic, $\dot{\theta} \equiv 0$ by our choice of
equatorial plane.  Finally, the radial equation of motion follows from
the fact that the 4-velocity $u^{\mu}$ of \OR{} particles is
null. Using (\ref{eq:tdoteom}) and (\ref{eq:phidoteom}) to eliminate
$\dot{t}$ and $\dot{\phi}$, we get
\begin{align}
  \label{eq:rdoteom}
  0 = g_{\mu\nu} u^{\mu}u^{\nu}
  &=
    -A \dot{t}^{2} + B \dot{r}^{2} + r^{2}\dot{\theta}^{2} +
    r^{2}\sin^{2}(\theta\equiv\frac{\pi}{2})\dot{\phi}^{2} \\
  &= -\frac{1}{A} + B \dot{r}^{2} + \frac{\impact^{2}}{r^{2}} \\
  \longrightarrow
  \dot{r}^{2}
  &=
    \frac{1}{AB}
    \left(
    1 - \frac{A\impact^{2}}{r^{2}}
    \right) \\
  \longrightarrow
  \dot{r}
  &=
    \pm
    \frac{1}{\sqrt{AB}}
    \left(
    1 - \frac{Ab^{2}}{r^{2}}
    \right)^{1/2}
    \, .
\end{align}

Thus the 4-velocity of our fiducial null geodesic is
\begin{eqnarray}
\label{eq:4velfiducial}
  u^{\mu}(t, r, \theta, \phi) =
  \left[
  \begin{array}{c}
    1/A \\
    \pm\sqrt{\frac{1}{AB}}
    \sqrt{1 - \frac{A\impact^2}{r^2}} \\
    0 \\
    \frac{\impact}{\sin(\theta)r^2}
  \end{array}
  \right]
  = \frac{1}{A}
  \left[
  \begin{array}{c}
    1 \\
    \pm\sqrt{\frac{A}{B}}
    \sqrt{1 - \frac{A\impact^2}{r^2}} \\
    0 \\
    \frac{A\impact}{r^{2}\sin(\theta)}
  \end{array}
  \right]
  \, .
\end{eqnarray}
More generally, if the spatial velocity of an \OR{} particle at $r$
makes an angle $\gamma$ when projected onto the sphere of radius $r$, then its
4-velocity is (where $\gamma = 0$ would be latitudinal and
$\gamma = \pi/2$ would be longitudinal)
\begin{eqnarray}
\label{eq:4velgeneric}
  u^{\mu}(t, r, \theta, \phi; \gamma) =
  \frac{1}{A}
  \left[
  \begin{array}{c}
    1 \\
    \pm\sqrt{\frac{A}{B}}
    \sqrt{1 - \frac{A\impact^2}{r^2}} \\
    \sin(\gamma)\frac{A\impact}{r^2} \\
    \cos(\gamma)\frac{A\impact}{\sin(\theta)r^2}
  \end{array}
  \right]
  =
  \frac{1}{A}
  \left[
  \begin{array}{c}
    1 \\
    {} \\
    \vec{v} \\
    {}
  \end{array}
  \right]
  \, ,
\end{eqnarray}
with $\vec{v}$ denoting the ordinary spatial velocity 3-vector in the
standard Schwarzschild coordinate basis.

%\subsection{Deriving the stress-energy tensor $T$}

% We can generate the stress-energy tensor via two conservation
% properties (angular momentum and energy), and one symmetry condition
% (rotational invariance).  A traditional short hand for the tangential
% directions is $d\Omega^2 = d\theta^2 + \sin^2(\theta) d\phi^2$.
% Conservation of angular momentum allows us to define the impact
% parameter $\impact$ of null particles via the following equation:
% \begin{eqnarray}
% \frac{dr^2}{d\Omega^2}  =  \frac{r^4}{B}\left(\frac{1}{\impact^2 A} -
% \frac{1}{r^2}\right).
% \end{eqnarray}
% It is a conserved quantity for null particles.  If we write this as a
% 4-vector say,
% \begin{eqnarray}
% v(r,\theta,\phi,\gamma) \equiv \left[ \begin{array}{c}
% 1/A \\
% \frac{\sqrt{1/A -\impact^2/r^2}}{\sqrt{B}} \\
% \sin(\gamma)\frac{\impact}{r^2} \\
% \cos(\gamma)\frac{\impact}{\sin(\theta)r^2}
% \end{array}
% \right]
% \end{eqnarray}
% where $\gamma$ is the angle that the null particle is traveling in if
% we project it onto the sphere.  

\subsection{\OR{}s as anisotropic fluids}
\label{sec:anisotropicfluid}

The stress-energy tensor for the entire \OR{} should be the sum of the
individual stress-energy tensors of all the constituent null
particles.  $T^{\mu\nu}$ for a single null particle with 4-velocity
$u$ is essentially that of a null dust
\begin{equation}
  \label{eq:nullparticleTmunuapprox}
  T \sim \rho u \otimes u
\end{equation}
with delta functions as appropriate in the energy density $\rho$.  

Consider a local orthonormal frame field with basis vectors and basis
one-forms
\begin{align}
  \label{eq:orthonormalframebasisvecs}
  e_{\hat{t}} &= \frac{1}{\sqrt{A(r)}}\partial_{t}
  &\quad
    de^{\hat{t}} &= -\sqrt{A(r)}dt \notag \\
  e_{\hat{r}} &= \frac{1}{\sqrt{B(r)}}\partial_{r}
  &\quad
    de^{\hat{r}} &= \sqrt{B(r)}dr \notag \\
  e_{\hat{\theta}} &= \frac{1}{r}\partial_{\theta}
  &\quad
    de^{\hat{\theta}} &= r d\theta \notag \\
  e_{\hat{\phi}} &= \frac{1}{r \sin\theta}\partial_{\phi}
  &\quad
    de^{\hat{\phi}} &= r \sin\theta d\phi
  \,.
\end{align}
Since all the particles have the same impact parameter
magnitude $\abs{\impact}$, the $\vec{v}$ distribution in a local
orthonormal frame is anisotropic. The projection $v^{\perp}$ of
$\vec{v}$ into the spatial plane normal to $e_{\hat{r}}$ is
isotropic within that plane, so that
\begin{equation}
  \label{eq:rms_vperp}
  \langle (v^{\hat{\theta}})^{2} \rangle = \langle (v^{\hat{\phi}})^{2} \rangle
  = \frac{1}{2} \langle (v^{\perp})^{2} \rangle
  \, .
\end{equation}
But, as we can see from (\ref{eq:4velgeneric}) and
(\ref{eq:orthonormalframebasisvecs}), the angle between $v^{\hat{r}}$
and $v^{\perp}$ varies with $r$, so that in general,
\begin{equation}
  \label{eq:vr_neq_vperp}
  \langle (v^{\hat{r}})^{2} \rangle \equiv
  1 - \langle (v^{\perp})^{2} \rangle
  \neq
  \frac{1}{2} \langle (v^{\perp})^{2} \rangle
  \, .
\end{equation}
Thus, the radial pressure
\begin{equation}
  \label{eq:Prdefn}
  P_{r} = \rho \langle (v^{\hat{r}})^{2} \rangle
\end{equation}
and the tangential pressure
\begin{equation}
  \label{eq:Pperpdefn}
  P_{\theta} = P_{\phi} =
  \frac{1}{2} \rho \langle (v^{\perp})^{2} \rangle
  \equiv P_{\perp}
\end{equation}
are unequal in general.

Taking the continuum limit, an \OR\ is an anistropic fluid whose
stress-energy tensor in the local orthonormal rest frame of a fluid
element is
\begin{equation}
  \label{eq:Tmunuanisotropicinrestframe}
  T^{\hat{\mu}\hat{\nu}} =
  \text{diag} (\rho, P_{r}, P_{\perp}, P_{\perp})
  \, ,
\end{equation}
or, in manifestly covariant form (see, e.g., \cite{Anisotropic2007} or
\cite{Gergeley}),
\begin{equation}
  \label{eq:Tmunuanisotropicfluid}
  T^{\mu\nu} = (\rho + P_{\perp}) u^{\mu}u^{\nu} + P_{\perp}g^{\mu\nu}
  + (P_{r} - P_{\perp}) e_{\hat{r}}^{\mu}e_{\hat{r}}^{\nu}
  \, .
\end{equation}

Even though the fluid consists of null particles, each resulting fluid
element has a timelike 4-velocity $u^{\mu}$ and a well-defined rest
frame. However, because the \OR{} consists of null particles, we still
expect this $T^{\mu\nu}$ to be traceless.  And it is -- the speed of
light is unity in all local Lorentz frames, so from (\ref{eq:Prdefn})
and (\ref{eq:Pperpdefn}, we get
\begin{equation}
  \label{eq:anisotropic_fluid_eos}
  P_{r} + 2 P_{\perp} =
  \rho (\langle (v^{r})^{2} \rangle +
  \langle (v^{\perp})^{2} \rangle
  = \rho
  \, .
\end{equation}

\subsection{Explicit expression for $T_{\mu\nu}$}
\label{sec:Tmunuexplicit}

To conclude this section, we derive an explicit expression for the
stress-energy tensor (\ref{eq:Tmunuanisotropicfluid}) in terms of $r$,
the (still unknown) metric coefficient functions $A(r)$ and $B(r)$,
and a small number of parameters that characterize an \OR{}.

Let $n_{+}(r)$ and $n_{-}(r)$ denote the number densities at $r$ (as
measured in frame (\ref{eq:orthonormalframebasisvecs})) of null
particles with $v^{\hat{r}} > 0$ (outward bound) and $v^{\hat{r}} < 0$
(inward bound), respectively. $n_{+}(r)$ must equal $n_{-}(r)$ at
every $r$ -- if, say, $n_{+}$ were larger, then when the radial
velocities of all the particles changed sign after half a radial
period of the null geodesics, $n_{-}$ would be larger, counter to the
assumption of staticity.

As measured in frame (\ref{eq:orthonormalframebasisvecs}), the radial
component of the spatial velocity $\vec{v}$ for null particles moving
radially outward ($v^{\hat{r}} > 0$)across the spherical surface at
$r$ is (using (\ref{eq:4velgeneric}) and
(\ref{eq:orthonormalframebasisvecs}))
\begin{align}
  \label{eq:vlocintermsofr}
  v^{\hat{r}} \equiv d\hat{r}/d\hat{t}
  &= \frac{\sqrt{B}}{\sqrt{A}} \frac{dr}{dt} \\
  &= \frac{\sqrt{B}}{\sqrt{A}} \frac{\sqrt{A}}{\sqrt{B}}
    \sqrt{1 - \frac{A\impact^2}{r^2}} \\
  &= \sqrt{1 - \frac{A\impact^2}{r^2}}
    \, .
\end{align}
In the local orthonormal frame (\ref{eq:orthonormalframebasisvecs}),
these radially outward-bound particles cross the spherical surface at
$r$ at a rate
\begin{align}
  \label{eq:dN+ofr_tloc}
  \frac{dN}{d\hat{t}}
  &= n_{+}(r) 4\pi r^{2} v^{\hat{r}} \\
  &= n_{+}(r) 4\pi r^{2} \sqrt{1 - \frac{A\impact^2}{r^2}}
  \, ,
\end{align}
where $n_{+}(r)$ denotes the (also locally measured) number density of
outward flowing particles. Because the system must be time-reversal
invariant, we must have
\begin{equation}
  \label{eq:npluseqnminus}
  n_{+}(r) \equiv n_{-}(r) \equiv \frac{1}{2} n(r)
  \, ,
\end{equation}
where $n(r)$ is the total local particle number density (outbound and
inbound).

In terms of the Schwarzschild time coordinate $t$, the rate
corresponding to (\ref{eq:dN+ofr_tloc}) is time-dilated to
\begin{align}
  \label{eq:dN+ofr_tinf}
  \frac{dN_{+}}{dt}
  &= \sqrt{A} \frac{dN_{+}}{d\hat{t}} \\
  &= \sqrt{A} n_{+}(r) 4\pi r^{2} v^{\hat{r}} \\
  &= \sqrt{A} \frac{1}{2} n(r) 4\pi r^{2} \sqrt{1 - \frac{A\impact^2}{r^2}}
    \, .
\end{align}
By staticity, this rate must be independent of $t$.  It must also be
independent of $r$ -- after one coordinate radial period $T_{r}$ of
the geodesics, every particle returns with the same velocity to
the spherical surface from which it departed, and the ``geodesic
conveyor belt'' has carried every particle in the \OR{} outward
across the spherical surface at $r$ exactly once. Therefore, if the
total number of particles in the \OR{} is $N$, then
\begin{equation}
  \label{eq:fluxisconst}
  2\pi \sqrt{A(r)} n(r) r^{2} \sqrt{1 - \frac{A(r)\impact^2}{r^2}}
  \equiv \frac{N}{T_{r}}
  \, ,
\end{equation}
and the local particle number density is
\begin{equation}
  \label{eq:nofrexpression}
  n(r) =
  \frac{N/T_{r}}{2\pi}
  \frac{1}{\sqrt{A(r)} r^{2} \sqrt{1 - \frac{A(r)\impact^2}{r^2}}}
  \, .
\end{equation}

Now multiply both sides of (\ref{eq:nofrexpression}) by the average
particle energy in the local frame
$\langle \Eloc \rangle = \langle \Einf \rangle / \sqrt{A(r)}$. On the
LHS, we get the local energy density
\begin{equation}
  \label{eq:rhoeqntimesEloc}
  \rho(r) \equiv n(r) \langle \Eloc \rangle(r)
\end{equation}
that appears in (\ref{eq:Tmunuanisotropicfluid}). On the RHS, we get
\begin{align}
  \label{eq:RHSofrhoofr}
  &\phantom{=}\frac{\langle \Einf \rangle N/T_{r}}{2\pi}
    \frac{1}{\sqrt{A(r)}}
    \frac{1}{\sqrt{A(r)} r^{2} \sqrt{1 - \frac{A(r)\impact^2}{r^2}}}\\
  &=\frac{M_{\text{OR}}/T_{r}}{2\pi}
    \frac{1}{A(r) r^{2} \sqrt{1 - \frac{A(r)\impact^2}{r^2}}}
    \, ,
\end{align}
where $M_{\text{OR}}$ is the total mass of the \OR{} measured by
distant observers, and $M_{\text{OR}}/T_{r}$ is the ``energy
circulation rate'' of particles within the \OR{}.

We absorb all the prefactors into a single constant
\begin{equation}
  \label{eq:fluxconst_defn}
  \flux \equiv \frac{M_{\text{OR}}/T_{r}}{2\pi}
\end{equation}
and define a convenience function
\begin{equation}
  \label{eq:m_of_r_defn}
  m(r) \equiv \frac{\flux}{r^{2} \sqrt{1 - b^{2} A(r)/r^{2}}}
  \, .
\end{equation}
The local energy density in the \OR{} is then
\begin{equation}
  \label{eq:rho_of_r_final}
  \rho(r) = \frac{\flux}{A(r) r^{2} \sqrt{1 -
      \frac{A(r)\impact^2}{r^2}}}
  = \frac{m(r)}{A(r)}
  \, .
\end{equation}
Using (\ref{eq:rho_of_r_final}), (\ref{eq:Prdefn}),
(\ref{eq:Pperpdefn}) and (\ref{eq:4velgeneric}), the pressures
$P_{r}(r)$ and $P_{\perp}(r)$ can also be expressed in terms of
$r, A, B, m(r)$ and $\impact$.

Combining those results with the metric (\ref{eq:metric}) and equation
(\ref{eq:Tmunuanisotropicfluid}), we arrive at the following
expression for the nonvanishing components (in the Schwarzschild
coordinate basis) of the \OR{} stress-energy tensor
\begin{subequations}
  \label{eq:Tmunuexplicit_20tensor}
  \begin{eqnarray}
    T^{tt} & = & \m(r)/A^2 \label{eqn:T1}\\
    T^{rr} & = & \frac{\m(r)}{AB}(1 - A\impact^2 /r^2) \\
    T^{\theta\theta} & = & \m(r) \impact^2/(2r^4) \\
    T^{\phi\phi} & = &T^{\theta\theta} / \sin^2(\theta) \label{eqn:T2}
                       \, .
  \end{eqnarray}
\end{subequations}
The $(0,2)$ version of the tensor
$T_{\mu\nu} = g_{\alpha\mu} g_{\beta\nu} T^{\alpha\beta}$ is
\begin{subequations}
  \label{eq:Tmunuexplicit_02tensor}
  \begin{eqnarray}
    T_{tt} & = & \m(r) \label{eqn:T1covariant}\\
    T_{rr} & = & \frac{\m(r)B}{A}(1 -  A\impact^2 /r^2) \\
    T_{\theta\theta} & = & \m(r) \impact^2/2 \\
    T_{\phi\phi} & = &T_{\theta\theta} \sin^2(\theta) \label{eqn:T2covariant}
                       \, .
  \end{eqnarray}
\end{subequations}

Note that an \OR{} is specified by two parameters in terms of which
all other parameters are determined: $\flux$ (the energy circulation
rate in the \OR{}, which we will sometimes refer to simply as ``the
flux''), and $\impact^{2}$ (the square of the impact parameter). The
total mass of the \OR{} is also a parameter, but adjusting it merely
rescales any given solution (see section~\ref{sec:family}).

% will
% the product of the is an average over particles according to or with
% all the dependencies
% \begin{eqnarray}
% T^{\alpha\beta} = m(r)
%  E(v(r,\theta,\phi,\gamma)^\alpha v(r,\theta,\phi,\gamma)^\beta)
% \end{eqnarray}
%  where $m(r)$ is the density of null particles at $r$.  Our spherically
%  symmetric model then tells us that $\gamma \sim U([0,2\pi])$, and so
%  the expected value of $\sin(\gamma)^2$ is $1/2$.  So the stress-energy
%  tensor imposed by the average of particles is:
% \begin{eqnarray}
% T^{tt} & = & \m(r)/A^2 \label{eqn:T1}\\
% T^{rr} & = & \frac{\m(r)}{B}(1/A -  \impact^2 /r^2) \\
% T^{\theta\theta} & = & \m(r) \impact^2/(2r^4) \\
% T^{\phi\phi} & = &T^{\theta\theta} / \sin^2(\theta) \label{eqn:T2}
% \end{eqnarray}
% We seek and verify a static solution so these equations are time
% independent.  Using conservation of energy, we get
% \begin{equation}
% \label{eq:m_of_r_defn}
% \m(r) = \flux / (r^2\sqrt{1 - \impact^2 A/r^2})
% \end{equation}
% for a constant $\flux$.

\section{The Einstein equations for an \OR{}}
\label{sec:einsteineqns}

\subsection{Derivation of ODEs}

As discussed in Section \ref{sec:einsteintensor}, the Einstein
equations for an \OR{} become
\begin{equation}
  \label{eq:Rmunu=Tmunu}
  R_{\mu\nu} = 8\pi G T_{\mu\nu} \,,
\end{equation}
with the left- and right-hand sides, respectively, defined by
(\ref{eq:Riccitensor}) and (\ref{eq:Tmunuexplicit_02tensor}).  As we
demonstrate in the Appendix, the fact that
\begin{equation}
  \label{eq:Tmunutraceless}
  \tr T \equiv T^{\mu}_{\phantom{\mu}\mu} = 0 \, 
\end{equation}
helps us reduce \eqref{eq:Rmunu=Tmunu} to a pair of coupled, nonlinear
1st-order ODEs for the metric coefficients $A(r)$ and $B(r)$,
\begin{subequations}
  \label{eq:deanODEs}
  \begin{eqnarray}
    r A'/A & = &  \frac{\flux B \sqrt{1 -  \impact^2A/r^2}}{A} +
                 B - 1 \label{eqn:dean:A} \\
    r B'/B & = &   \frac{\flux B}{A\sqrt{1 - \impact^2A/r^2}} - B + 1
                 \label{eqn:dean:B}
                 \,,
  \end{eqnarray}
\end{subequations}
where a prime denotes differentiation with respect to $r$.

To prove the above, we develop an equation for $rA'/A$ and
 $rB'/B$ which holds for all null particle problems.  In other words,
 only under the assumption that $T = g_{\alpha\alpha}T^{\alpha\alpha}
 = 0$, we can get some simplifications of the Einstein equations.  In
 particular, we eliminate the $A''$ from our equations
 for $R_{tt}$ and $R_{rr}$.

\begin{lemma}\label{lemma:SDT}
If $g_{\nu\nu}T^{\nu\nu} = 0$, we can
write our ODE in terms of $T$'s:
\begin{eqnarray}
A'/A & = &  - 16 \pi G Br^3 T^{\theta\theta} + 8 \pi G rBA T^{tt} +
B/r - 1/r \\
B'/B & = &  8 \pi G rBA T^{tt} - B/r + 1/r
\end{eqnarray}
\end{lemma}

The proof is in the appendix.

{\bf Proof of equations (\ref{eqn:dean:A}) and (\ref{eqn:dean:B}):}
From equations \ref{eqn:T1},\ref{eqn:T2}, and (\ref{eq:m_of_r_defn}) we get:
\begin{eqnarray*}
T^{tt} & = & \m/A^2 \\
T^{\theta\theta} & = & \m \impact^2/(2r^4)
\end{eqnarray*}
Plugging these in to lemma \ref{lemma:SDT} we get:
\begin{eqnarray*}
A'/A & = &  - 16 \pi G Br^3 \m \impact^2/(2r^4) + 8 \pi G rBA \m/A^2 +
B/r - 1/r \\
B'/B & = &  8 \pi G rBA \m/A^2 - B/r + 1/r
\end{eqnarray*}
which simplifies as:
\begin{eqnarray*}
A'/A & = &  - 8 \pi G B \m \impact^2/r + 8 \pi G rB \m / A +
B/r - 1/r \\
B'/B & = &  8 \pi G rB \m/A - B/r + 1/r
\end{eqnarray*}
Now plugging in $\m$ from equation \ref{eq:m_of_r_defn} and setting
$\ALTdensity = 8 \pi G \flux$ we get:
\begin{eqnarray*}
A'/A & = &  \ALTdensity \frac{rB/A - B  \impact^2/r}{r^2\sqrt{1 - \impact^2A/r^2}} +
B/r - 1/r \\
B'/B & = &   \frac{\ALTdensity B}{rA\sqrt{1 - \impact^2A/r^2}} - B/r + 1/r
\end{eqnarray*}
which is
\begin{eqnarray*}
A'/A & = &  \ALTdensity (Br/A) \frac{1 -  \impact^2A/r^2}{r^2\sqrt{1 - \impact^2A/r^2}} +
B/r - 1/r \\
B'/B & = &   \frac{\ALTdensity B}{rA\sqrt{1 - \impact^2A/r^2}} - B/r + 1/r
\end{eqnarray*}
and simplifies to
\begin{eqnarray*}
A'/A & = &  \frac{\ALTdensity B \sqrt{1 -  \impact^2A/r^2}}{Ar} +
B/r - 1/r \\
B'/B & = &   \frac{\ALTdensity B}{rA\sqrt{1 - \impact^2A/r^2}} - B/r + 1/r .
\end{eqnarray*}
Multiplying by $r$ completes the proof.
\hfill$\Box$

\subsection{The solution family}
\label{sec:family}

All \OR\ solutions have an inner radius and an outer radius.  The
inner radius $r_{\text{inner}}$ is the smallest radius that
constituent null particles achieve while the outer radius
$r_{\text{outer}}$ is the largest radius that constituent null
particles achieve.

Other key parameters are the flux $\flux$, the impact parameter
$\impact$, and the boundary conditions for $A$ and $B$.  Using
symmetries and dependencies, we reduce this to a one-dimensional
family of solutions.  To see this, we first describe two invariances
of the ODEs.

\begin{enumerate}
\item If $A$, and $B$ are solutions for (\ref{eqn:dean:A}) and
(\ref{eqn:dean:B}) with parameters $\impact$ and $\flux$, then for any constant $k$, $kA$
and $B$ are solutions for $\impact/\sqrt{k}$ and $k\flux$.
\begin{equation}
\label{eq:inv:A}
A(r), B(r) \hbox{ satisfy (\ref{eqn:dean:A})-(\ref{eqn:dean:B})}
\Longrightarrow
kA(r), B(r) \hbox{ satisfy (\ref{eqn:dean:A})-(\ref{eqn:dean:B})}
\end{equation}
We use this property in our simulations to set $A(r_{\text{inner}}) =
1$.  After the fact, this symmetry can be used to set $A(\infty) = 1$
as per convention.

\item
If $A(r)$ and $B(r)$ are solutions (\ref{eqn:dean:A}) and
(\ref{eqn:dean:B}) with parameters $\impact$ and $\flux$, then for any constant $k$,
$A(kr)$ and $B(kr)$ are solutions for $k\impact$ and $k\flux$.
\begin{equation}
\label{eq:inv:r}
A(r), B(r) \hbox{ satisfy (\ref{eqn:dean:A})-(\ref{eqn:dean:B})}
\Longrightarrow
A(kr), B(kr) \hbox{ satisfy (\ref{eqn:dean:A})-(\ref{eqn:dean:B})}
\end{equation}
This scale invariance implies we can set $r_{\text{inner}} = 1$ in our simulations.
\end{enumerate}

A few other simplifications are immediate.

\begin{enumerate}
\item We have $B(r_{\text{inner}}) = 1$ since we are seeking a
  solution which is flat inside the inner radius.   As long as the mass distribution is compact, this implies $B(\infty) =1$ as well.
\item Since $A(r_{\text{inner}}) = 1$ and $r_{\text{inner}} = 1$, we must have $b = 1$. 
\item Given $A(r_{\text{inner}}) = 1$, $B(r_{\text{inner}}) = 1$,
  $r_{\text{inner}} = 1$, and the flux $\flux$, the value of
  $r_{\text{outer}}$ is determined.
\end{enumerate}

The remaining free parameter is the flux $\flux$ which indexes the
family of solutions.

\section*{Semi-analytic solutions}

There are three special cases which can be approximated analytically.
The first is low flux.  Since low flux implies low curvature, photons
starting at the inner radius move almost directly away from the center
at large $r$.  So for large $r$, it is basically a counter flux of
inward photons and outward photons.  Hence the solution of Gergely
\cite{Gergeley} is a good approximation.  We can define $\flux_0$ as
the maximum flux for which a closed \OR\ is not formed.

A second analytic solution arises in the close-to-critical case.  Here
there is just barely enough intensity to close the ``orbits.''  In
this case, the photons end up spiraling to and away from the outer
radius with their radius changing infinitesimally, limiting to a
perfect circle.  The following theorem applies with proof in the Appendix.

\begin{theorem}\label{thm:three}  If $\impact = 1$ and
$\flux$ close to $\flux_0$, then $B(r) \approx 3$ and $A(r) \approx r^2$ is a solution
to our equations.  In particular, 
\begin{eqnarray}
A/r^2 & = & 1 - 9\flux^2 r^{-4}/4 + O(\flux^6r^{-12})\\
B & = & 3 + 9\flux^2 r^{-4}/2 + O(\flux^4r^{-8})
\end{eqnarray}
\end{theorem}

A third analytic solution arises in the high-flux case.  If the flux
is high enough, then the inner and outer radius are almost the same.
Hence there is basically no dependence on $r$ in the ODEs and they
can be solved exactly.  The following theorem applies with proof in
the Appendix.

\begin{theorem}\label{thm:nine} For high flux (i.e. $\flux \approx \infty$),  the
maximum achievable value for $B$ is $B\approx 9$, with
$r_{\text{inner}} \simeq r_{\text{outer}} \simeq 2.25$.
\end{theorem}

From these two theorems, we can catalog \OR s by the maximal value
that $B$ ever obtains.  If $B_{max} < 3$, then the \OR\ must be
open and the photons never return.  If $3 \le B_{max} \le 9$ then the
\OR\ is closed with a flat interior.  There are no OR-stars with a flat interior
  spacetime\footnote{We see later in Section \ref{sec:neg-mass} that
  $B_{\text{max}} > 9$ is possible with a negative mass singularity at
  the origin.} that have $B_{\text{max}} > 9$.

\section{Methods of Integration and Simulation}
\label{sec:simulation}
Given the partial differential equations from the previous section,
there are several ways to do an integration (or simulation) to
discover the geometry of the object.  We tried two in order to
convince ourselves that the results are sound.  Both approaches can be
easily implemented on a desktop computer and run fairly quickly.

\subsection{Integrating Differential Equations}

At $r_{\text{inner}}$ and $r_{\text{outer}}$, where the spatial
velocity of the null particles is instantaneously parallel to the
inner and outer spherical surfaces of the OR-star, the density
diverges to infinity like $\rho(r) \sim 1/\sqrt{r -
  r_{\text{inner/outer}}}$.  But this density still integrates to a
finite value, so that the total mass can be well-approximated
numerically by integrating very close to $r_{\text{inner}}$ and
$r_{\text{outer}}$.  We tested this approach using the deSolve package
in R which uses Runge-Kutta techniques for integration.

\subsection{Integration from the inner radius to the outer radius}

To confirm that we are not losing essential physics near the inner and
outer radius, we also tried an alternative numerical approach, using
as the integration variable the affine parameter of an element of null
particles traveling from $r_{\text{inner}}$ to $r_{\text{outer}}$.

In essence, this approach does integration-by-simulation.  Since
spherical symmetry exists, we can pick any element at the
inner radius and follow its trajectory to the outer radius, referring
to the element's radius to index a differential shell of elements of two sorts:
\begin{enumerate}
\item Elements starting simultaneously\footnote{Simultaneous is
  defined with respect to an observer at the origin.} on the inner
  radius with positions uniform on the inner sphere and trajectories
  drawn from uniform tangents to the inner sphere.
\item Elements of the same mass/energy at the same radius returning
  from a hypothesized outer radius.  Conservation principles imply
  that such elements must have the same mass/energy as elements of the
  first sort differing only in the sign of their radial velocity.
\end{enumerate}

This simulation is self-consistent if the curvature generated creates
an outer radius where the geodesic has no radial component.  At this
point, the simulation can halt because we have already accounted for
the contribution of elements returning from the outer radius to the
inner radius.

In more detail, we can start the simulation with $r=1, A(1)=1, B(1)=1$
at the inner radius and a free parameter given by $\flux$ as discussed
in section~\ref{sec:family}.  Furthermore, the null particle must be
tangential to the inner radius, implying $\impact = 1$.

If we let $\delta$ be the simulation scale, the equations for the
forward step then become (after some algebra):
\begin{enumerate}
\item $m \leftarrow r \frac{B-1}{2B}$ //the mass inside the radius
\item $dm \leftarrow \delta \frac{4 \pi \flux \sqrt{r - 2 m}}{ r^{1/2}A^{3/2}}$ //the change in mass
\item $\phi \leftarrow \frac{1}{2} \ln(A)$//the current angle of null particle
\item $d\phi \leftarrow \delta \frac{A m + 4 \pi \flux \sqrt{r^2 - A}}{r^3A^{3/2}} $//the change in angle
\item $dr \leftarrow \delta \frac{\sqrt{r^2 - A}}{r \sqrt{AB}}$//the change in radius
\end{enumerate}

This leads to new values according to:
\begin{enumerate}
\item $r \leftarrow r + dr$
\item $m \leftarrow m + dm$
\item $B \leftarrow \frac{r}{r - 2  m}$
\item $\phi \leftarrow \phi + d\phi$
\item $A \leftarrow e^{2 \phi}$
\end{enumerate}

We can record the values of $r,A,B$ and simulate forward to the outer
radius.  With a self-consistent solution for the shell, attaching a
Schwarzschild geometry onto the outer radius provides a solution
everywhere.  Then $A(r)$ can be rescaled so $A(r)=1$ in
the asymptotic rest frame.

We implemented this integration/simulator using C++ doing simple
first-order integration according to Euler's method (as above).  The
results are quantitatively stable to high precision and agree with the
Runge-Kutta approach.  Our code is available~\cite{code}.

\subsection{Ray Tracing}

In order to test our understanding of the OR-star object, we also
created a four-dimensional ray-tracer to probe the family of geodesics
that test-particle photons traverse \footnote{These photons are
  test-particles for observation only, distinct from the null particle
  constituents of the OR-star.  Thus, they need a high enough energy
  to have a small wavelength with respect to the scale of the OR-star
  but a small enough energy not to disturb the solution.}.  This
process is a simple computation given the geometry $A(r)$ and $B(r)$
and conservation of angular momentum.

To do this efficiently, we recorded the geometry at each discrete
radius given by the integration/simulation above and used a simple
linear interpolation between the two nearest recorded points.

Searching for the geometry can be done for a point at an arbitrary
radius by first testing whether the radius is within the shell of the
object.  If it's interior, the geometry is flat.  If it's exterior,
then the geometry is Schwarzschild.  If it's in the shell, then a
binary search suffices, although a hinted location reduces this search to
amortized constant time in the case of ray tracing.

\section{Simulation results}
\label{sec:results}
We exercised our simulations in various ways to both check correctness
and understand the implications of \OR s.  

An important concept is the characteristic geodesic.  A characteristic
geodesic is the path that null particles takes when starting tangential to
the inner radius at the inner radius.  When this is gravitationally
closed, the characteristic geodesic can be glued together from inner
radius to outer to inner to outer etc... to get the path of null particles.

\subsection{Varying initial conditions}

Here we exercise initial conditions to get a sense of the solution structure.

\subsubsection{Varying radius}

\begin{figure}
\includegraphics{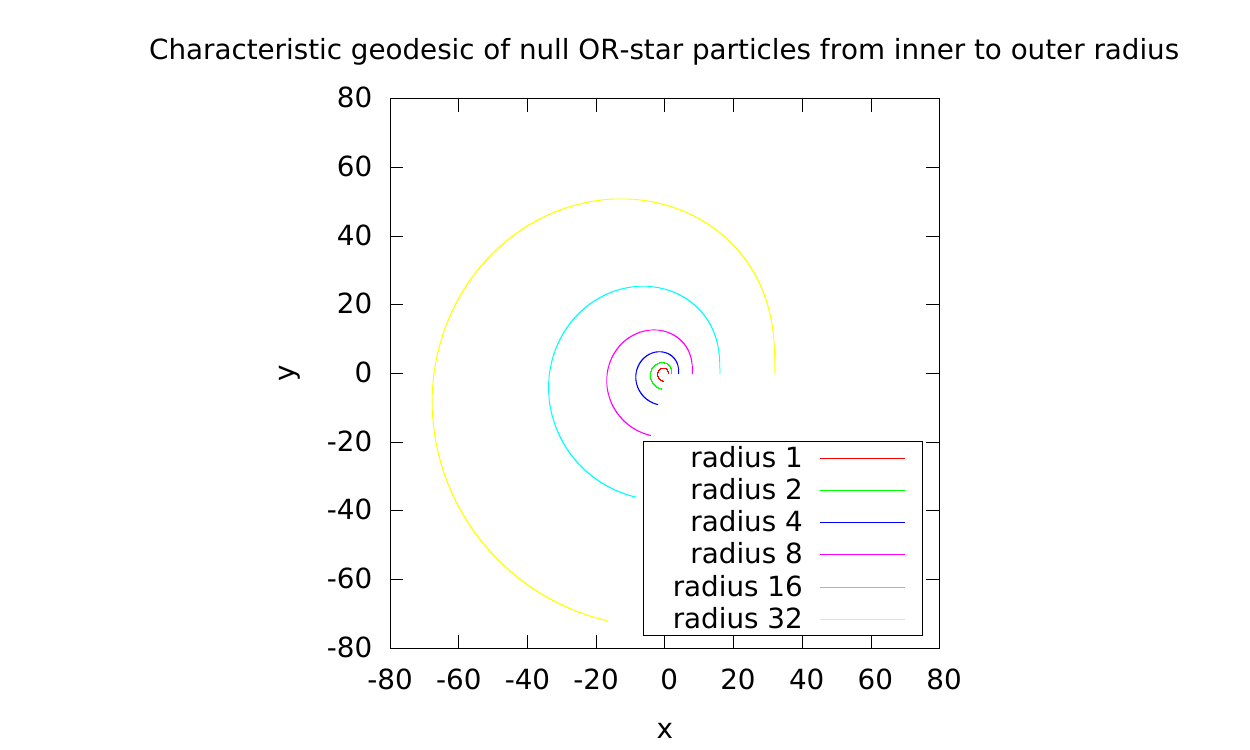}
\caption{A plot of the characteristic geodesic from inner to outer
radius for \OR s with varying radius but the same flux.  This graphically shows (\ref{eq:inv:r}).}
\label{fig:radius}
\end{figure}

Figure~\ref{fig:radius} shows the structure of the
characteristic geodesic is invariant to the radius, implying that the
scale of the structure can be arbitrary, as expected.

\subsubsection{Low mass/energy flux}

\begin{figure}
\includegraphics{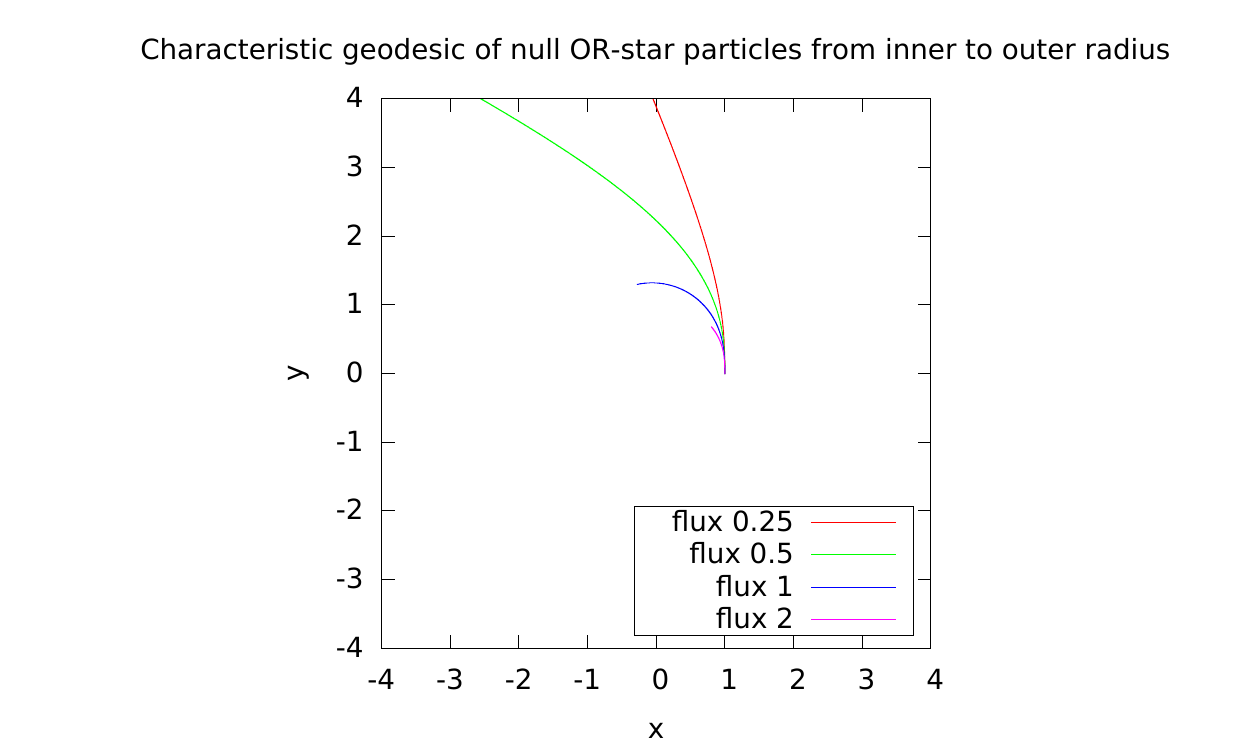}
\caption{The characteristic geodesic for \OR s with the same inner
  radius but different flux.}
\label{fig:flux}
\end{figure}

\begin{figure}[t!]
\centerline{\includegraphics[width=4in]{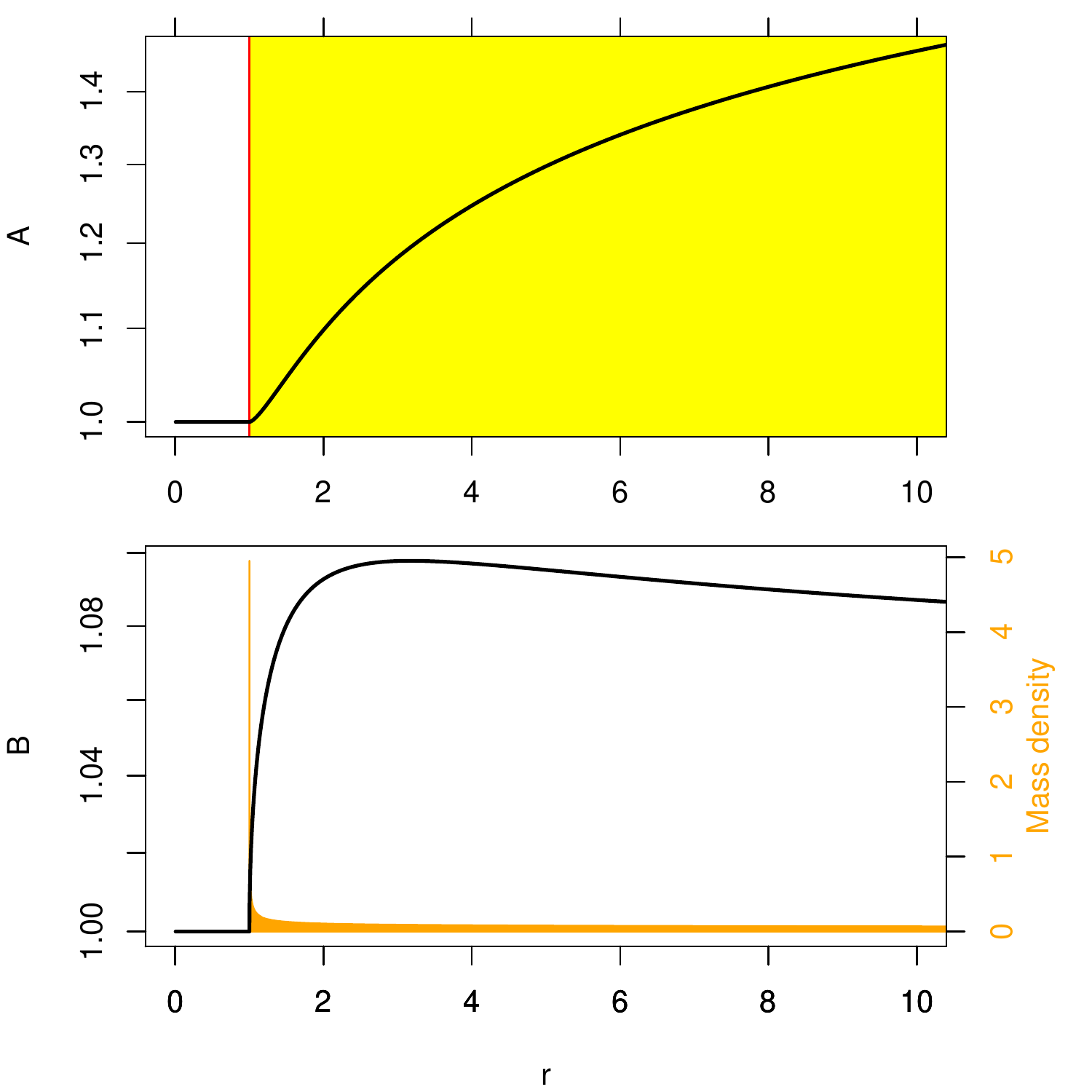}}
\caption{An ``open'' \OR\ with insufficient null particles for a compact
  object.  The upper plot shows $A$ with a yellow background showing
  the region of null particles.  The lower plot shows the value of
  $B$ and the density of the mass.}
\label{fig:open}
\end{figure}

In Figure~\ref{fig:flux} we see that the structure of the
characteristic geodesic as the mass/energy flux varies.  This result
has more structure, because the results with 0.25 or 0.5 flux are not
closed.

This result is as expected---low amounts of mass/energy should not be
able to generate sufficient curvature for geodesics to curl back.

In Figure~\ref{fig:open} we show the time geometry $A = g_{tt}$ and
radial geometry $B = g_{rr}$.  The difference between our model and
colliding null particles~\cite{Gergeley} is that colliding null particles head
directly towards or away from the center---an impact parameter of
zero.  Making this works requires a negative mass singularity at the
origin.  We avoid this with null particles having a greater impact
parameter, generating an inner radius of closest approach.  At a few
multiples of the inner radius from origin, the photons head almost
exactly away from the origin so the equations are similar to the
colliding null particles solution.

For such an open object, the maximal value of $B$ is always less than
3--namely the $B$ corresponding to the photon sphere.  The open object
with the largest $B$ has null particles orbiting in a photon sphere
at each possible radius.

\begin{figure}[t!]
\includegraphics{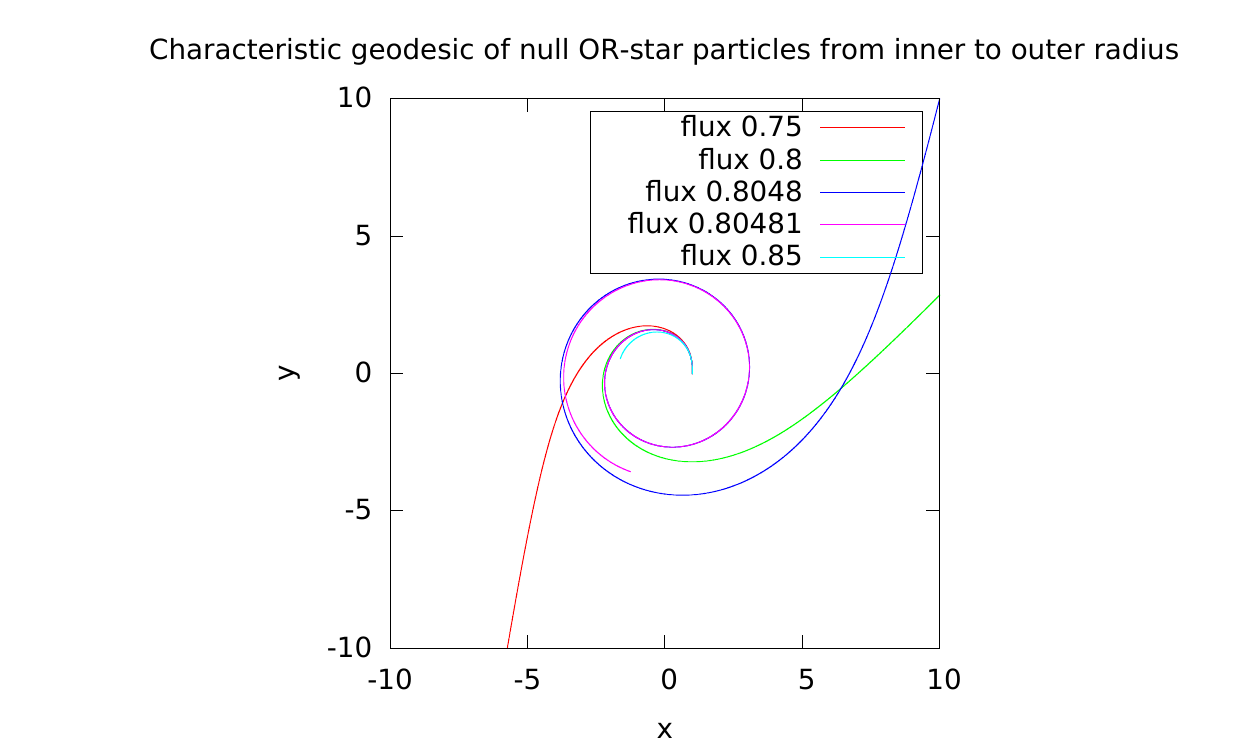}
\caption{A plot of the characteristic geodesic from inner to outer radius for \OR s with the same inner radius but varying flux near the critical flux.  Geodesics reaching the edge of the plot do not close.}
\label{fig:flux_critical}
\end{figure}

\subsubsection{The critical mass/energy flux}

In Figure~\ref{fig:flux_critical} we see variations around the
critical point where curvature becomes sufficient to create closed
geodesics.  In this plot, all non-closed geodesics reach the graph
edge and are valid when the null particles are reflected towards the origin.

\begin{figure}
\centerline{\includegraphics[width=4in]{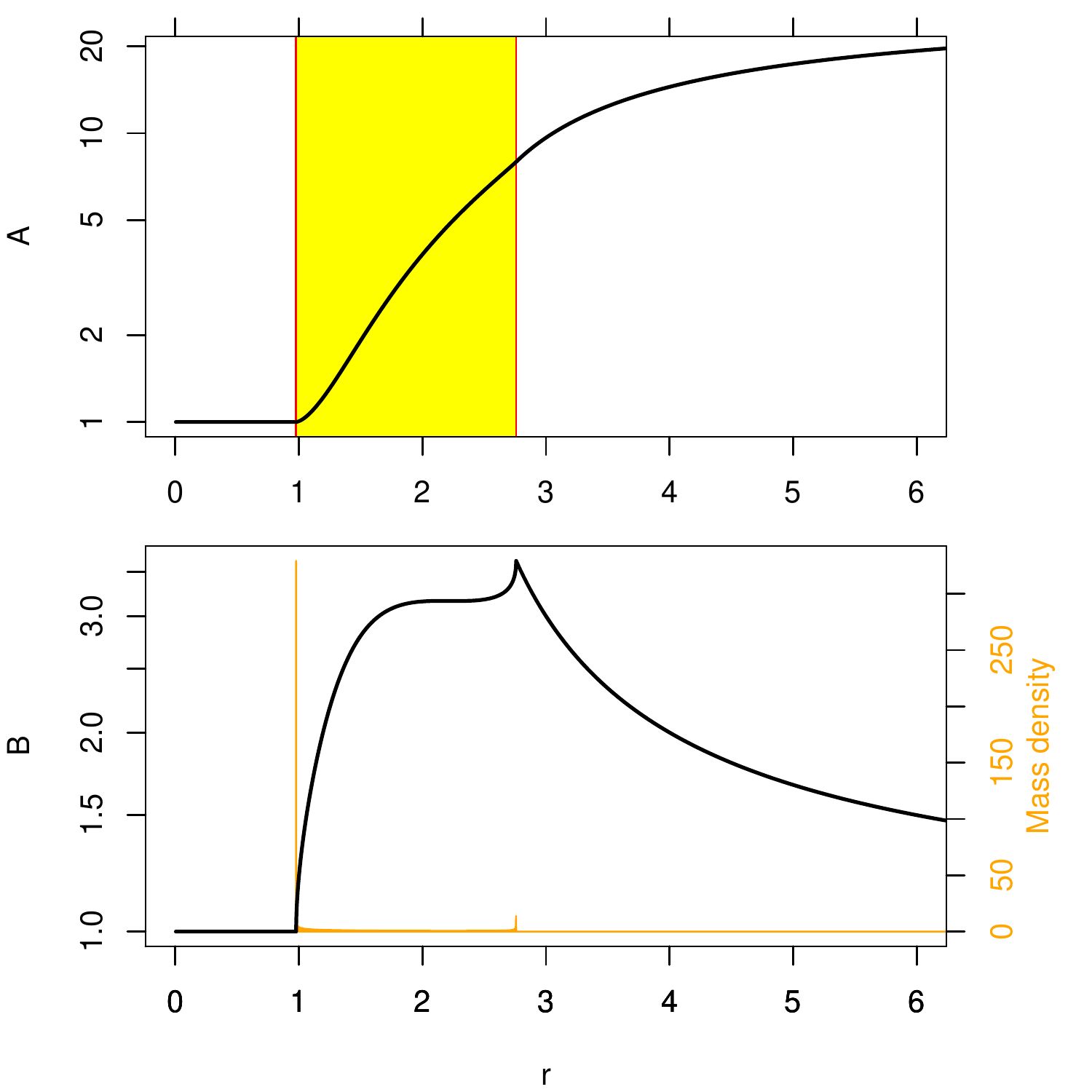}}
\caption{This is a near critical \OR.  The
  upper plot shows $A$ with a yellow background showing
  the region with null particles.  The lower plot shows the value of
  $B$ and the density.}
\label{fig:critical}
\end{figure}

Figure~\ref{fig:critical} shows $A$, $B$, and the potential well
created by a near-critical closed solution.  $B$ just barely increases
past $3$ and the null particles returns towards zero.  Near-critical
closed solutions are unstable since photons significantly off the
characteristic curve escape to infinity.

\subsubsection{High mass/energy flux}

The final example in this section is a high flux version (see figure
{\ref{fig:tightest}).  The maximal $B = 1/(1 - 2/2.25) = 2.25/.25 = 9$
  matches the bound in Theorem \ref{thm:nine} for the maximal mass which
  doesn't collapse into a black hole.  This is a much more stable
  object as illustrated by the trapped photons of
  Figure~\ref{fig:trapped} in the introduction.

\begin{figure}
\centerline{\includegraphics[width=4in]{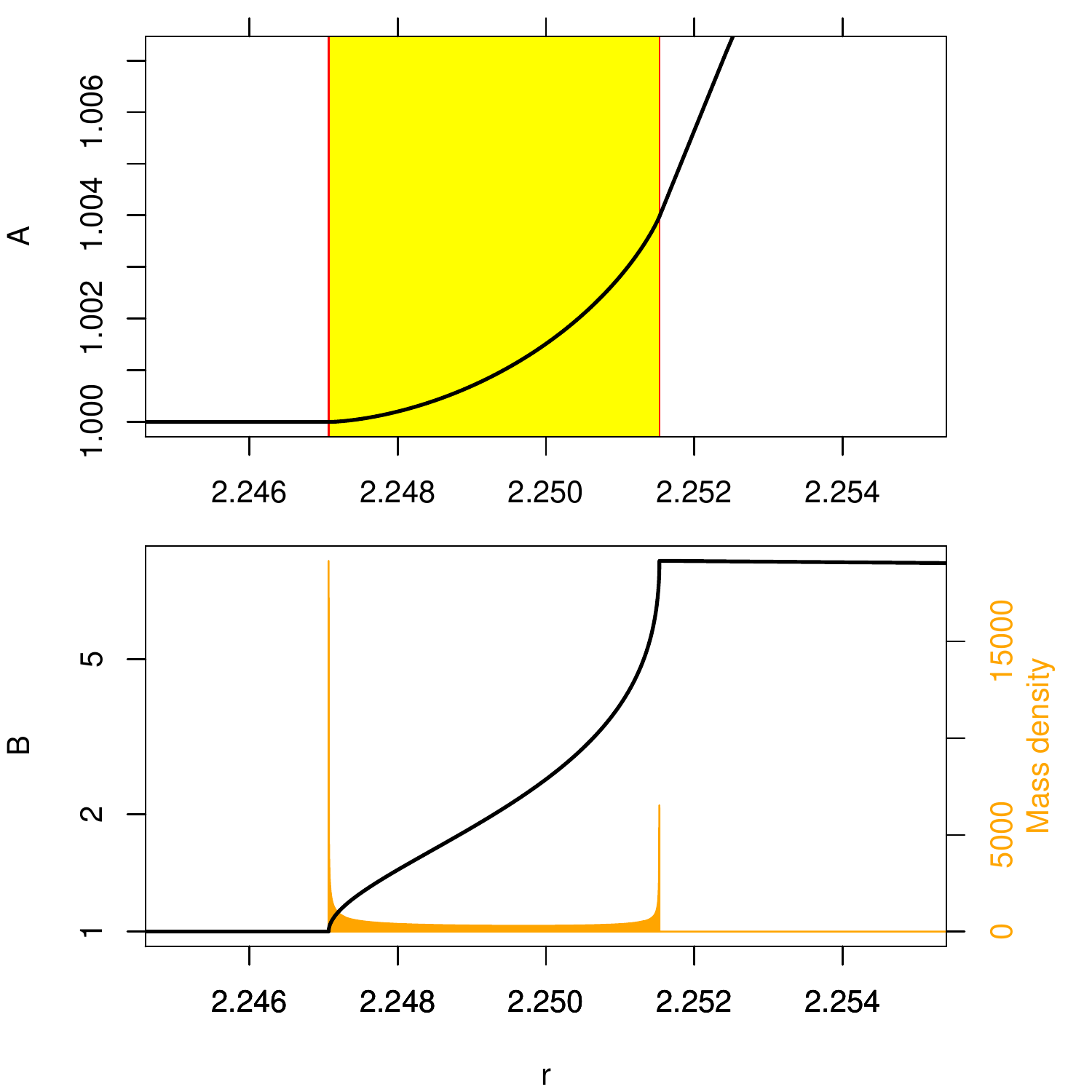}}
\caption{
This is a high flux \OR.  The upper plot shows $A$ with a yellow background showing the
  region with null particles.  The lower plot shows the value of $B$
  and the density.}
\label{fig:tightest}
\end{figure}

\subsection{Observational differences from a black hole}

\begin{figure}
\includegraphics{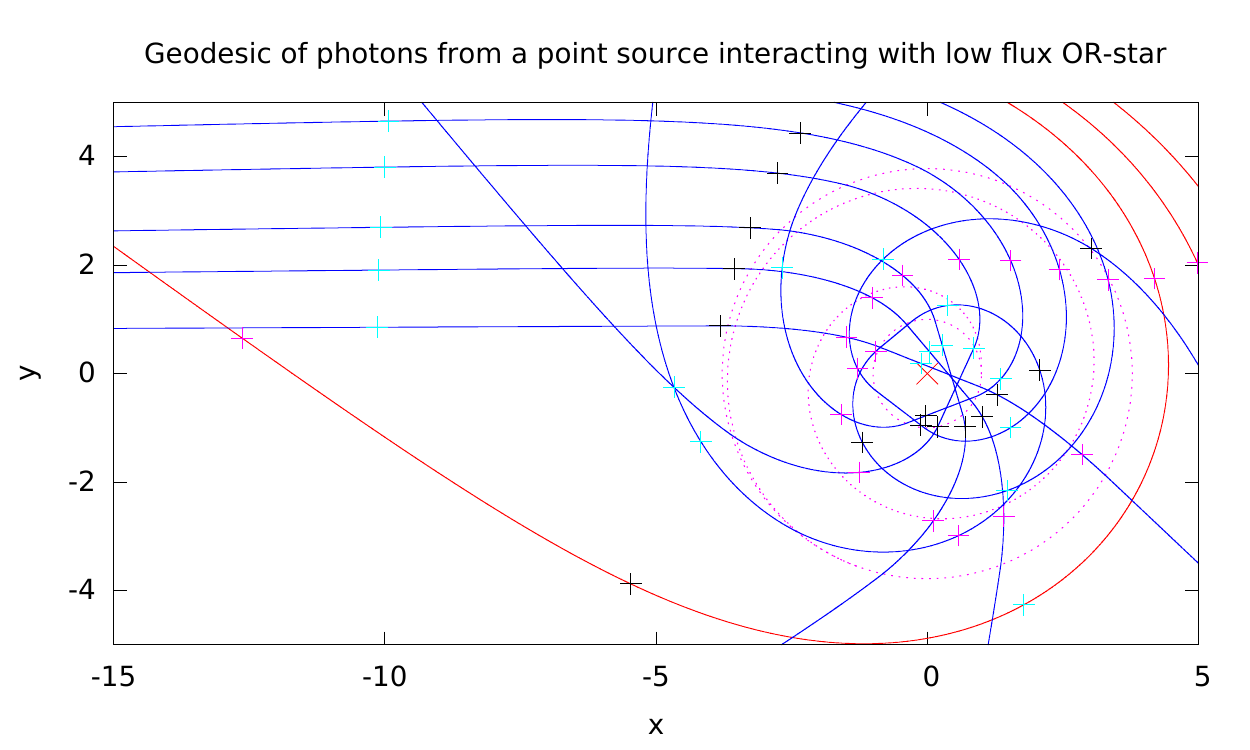}
\includegraphics{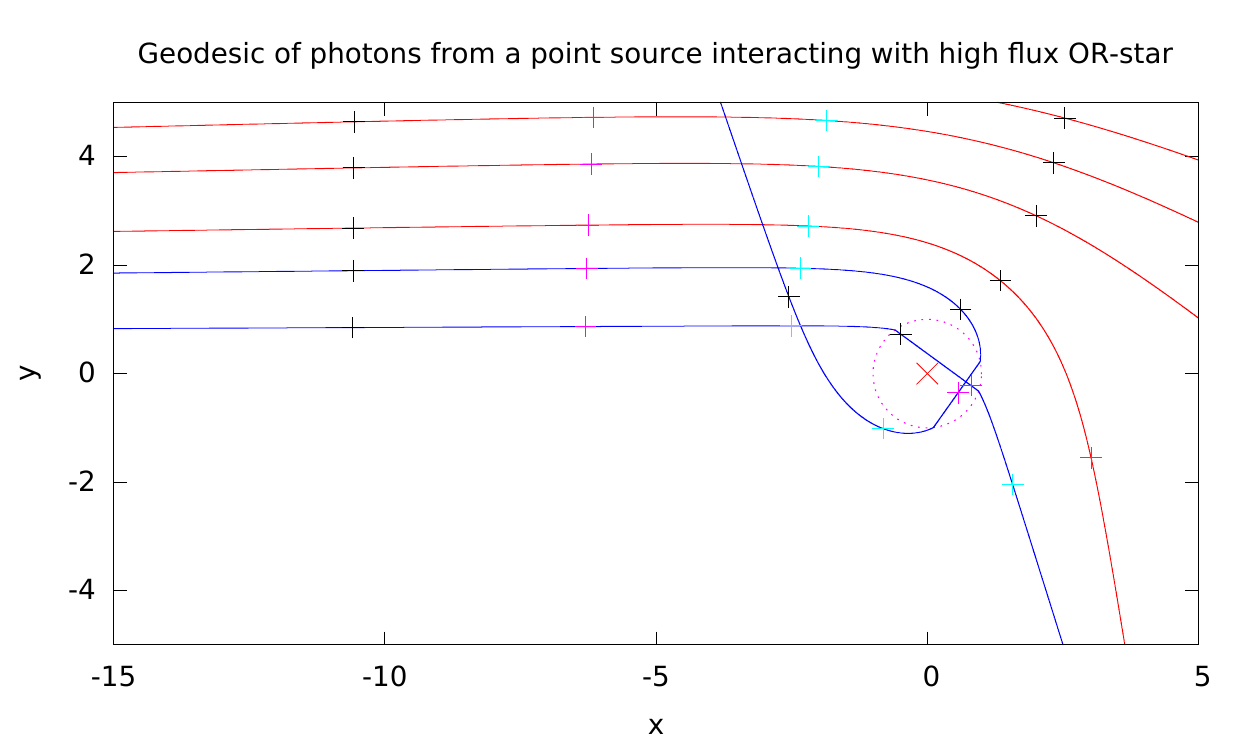}
\caption{Geodesics as traced by photons from a point source far to the
  left interacting with a near-critical \OR\ (above) and high flux
  \OR\ (below) at the origin.  The origin is marked in red, and the
  structure of the \OR\ (inner radius, outer radius, and
  characteristic geodesic) are shown with dotted lines.  For the high
  flux \OR\ these overlap leaving only a circle.  The black, red, and
  light blue tics mark equal separation in time as measured by the
  point source.  The red lines are Schwarzschild geodesics while the
  blue lines are not.}
\label{fig:probe}
\end{figure}

The structure of the object is similar to but observationally
different from a black hole because geodesics from infinity can probe
the geometry at all points and return to infinity.  In
Figure~\ref{fig:probe} we show the geodesics traced by photons from a
point source far away.  The expected external structure of
Schwarzschild geometry is observed, but the internal structure is
substantially more complicated.  It's easy to observe that the tics of
coordinate time advance more slowly in the interior from the viewpoint
of an observer far from the object, as expected.

\begin{figure}
\includegraphics{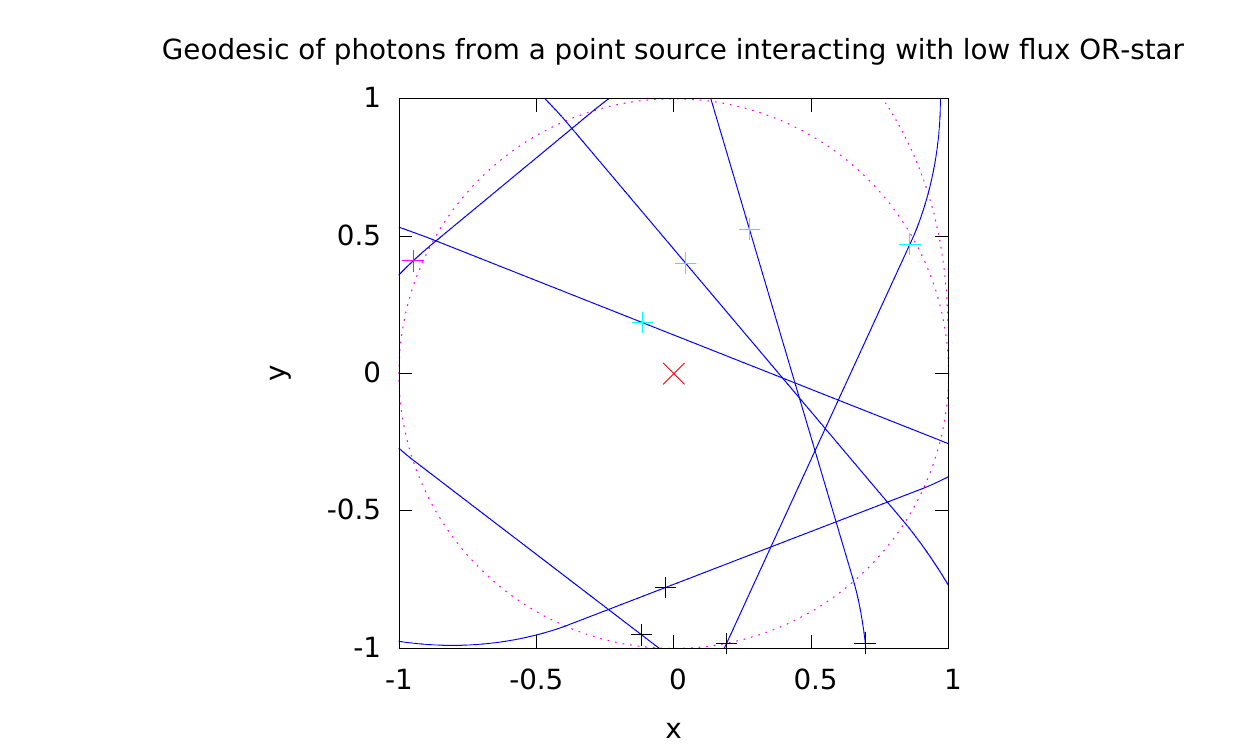}
\caption{A zoom on the same plot as figure~\ref{fig:probe}
  showing that the focal point is imperfect and the internal space is
  flat as expected.}
\label{fig:probe_double_zoom}
\end{figure}

In Figure~\ref{fig:probe_double_zoom} we zoom in to see the internal
structure in the flat internal region.  As expected, the geodesics are
all straight lines in this region, although it is interesting to
notice that the geodesics all focus on a point.  The focal point is
well within the inner radius and clearly imperfect.

Considering the global structure again, the object differs from a
black hole observationally for any geodesic crossing the outer radius.
Every photon crossing the outer radius is less tangential than the
null particles composing the \OR.  This monotonicity property is
preserved as the photon and null particles travel to the inner radius where
the photon must therefore enter the inner radius since it is not
tangential.  

Photons aimed directly at the center have their coordinate time
retarded by the mass of the object but are otherwise unaltered as
observed asymptotically.  Hence, the object acts as a ``timescope''
allowing an observer to record what happened on the other side of the
object in the past.

One observational signature of the \OR\ for a stationary observer
is therefore a combination of normal gravitational lensing via
Schwarzschild geodesics plus a time delayed observation of a point
source.  When the point source and the observer are moving relative to
the \OR, it is still possible to observe a signature, because
there often exists a geodesic between the point source and the
observer passing through the object.  Observation of the geodesic
passing through the object is difficult because small changes in input
angle from the source result in large changes in the required angle of
an observer.  In essence, an \OR\  lights up for observers at many
angles with only a small solid angle of input photons.  Since the
small angle of input photons are shared by many observers, it is
relatively dim.

Hence, a second signature of an \OR\ is given by widely dispersed
observers that record a dim version of the same time-delayed event
nearby to the object.  On such possibility would be given by a nearby
star that novas.  First, observers would see nova via Schwarzschild
geodesics, and then a dim time-delayed ``echo'' of the nova event via
non-Schwarzschild geodesics passing through the \OR.

\section{Solution Variations }
\label{sec:variations}

Several variations on the object are straightforward given the above.
We discuss them here for further consideration.

\subsection{Mass inside the inner radius}
\label{sec:mass-inside}
Up to this point we have been using a flat geometry inside the inner
shell, but other spherically symmetric inner geometries can be handled
as long as geodesics starting on the inner sphere increase in radius.
The only modification to the inner solution is an increased amount of
time dilation which can be determined after the fact as per our
existing simulation/solution.  The calculation of the characteristic
geodesic and the geometry between the inner and outer shell is only
modified by a change to the initial conditions in the inside-outside
simulation approach.  In essence, instead of starting with a mass of
$0$, you start with the mass of the inner solution using the geometry
that this implies.

\subsection{An \OR\ with a (near) event horizon}
\label{sec:neg-mass}

If the inner contains flat space, then the smallest outer radius for a
mass 1 object is $2.25$ and the maximum value of $B$ is $9$.

But, if we allow more exotic geometries inside the inner radius then
the outer radius can be close to 2 with $B$ arbitrarily large, similar
to the event horizon of a black hole.  Consider two exotic geometries:
an anti-de Sitter and a negative mass singularity at the origin.  The
anti-de Sitter looks like:
\begin{displaymath}
ds^2 = -\left(1+ k r^2\right)dt^2 +
\left(1+k r^2 \right)^{-1}dr^2 + r^2 d\Omega_{n-2}^2
\end{displaymath}
and the negative mass of size $M_0$ looks like the following geometry
which has a singularity at zero:
\begin{displaymath}
ds^2 = -\left(1 + M_0/r\right)dt^2 +
\left(1+M_0/r\right)^{-1}dr^2 + r^2 d\Omega_{n-2}^2
\end{displaymath}

If we use either of these geometries for the interior, we can make the outer shell come
arbitrarily close to $2M$.  Handling this is
straightforward---we can simply set $B$ to have some smaller value for
the initial condition.

When the outer sphere and the inner sphere are both close to $2M$, we
have a thin shell of null particles dividing two vacuum regions similar to
an anti-de Sitter Gravastar~\cite{Grava2001}.  

In an \OR, the null particles eventually turn around and so we have a
Schwarzschild exterior.  We can make this turn around point occur as
close to $2M$ as is desired by placing a negative mass inside of the
inner shell.  This negative mass may be a singularity at the origin,
consist of negative mass with an internal volume, or fill the volume
of the inner sphere similar to an Anti-De Sitter space.

\subsection{A large inner radius}

There is no upper bound to the scale of the inner radius, so the only
practical upper bound is available mass-energy.  A natural question
therefore is: What would an observer inside the inner radius see?
\begin{enumerate}
\item The internal region has a bounded dimension with a definite
  border where the null particles have an integrable singularity.
  Interaction with the null particles at the border may be more or less
  easy to observe depending on the form of the null particles, but
  observations from beyond the border are certainly distinguishable
  due to blueshift/redshift effects from $A(r)$ varying.
\item Due to the internal focal point which an \OR\ creates, an
  internal observer could select different internal locations to gain
  a naturally magnified view of locations in the external universe.
\item Since the internal $A(r)$ is necessarily different from an
  external $A(r)$, all observations of external events are necessarily
  blueshifted relative to internally generated observations.
\end{enumerate}

\subsection{Varying particle energies}

The energy of the constituent null particles was never explicitly used
in our equations.  Consequently, null particles with varying energies
could be used to construct an \OR\ so long as the individual null
particle energies are a negligible fraction of the total mass $M_{\text{OR}}$ of the OR-star.
The negligible fraction constraint is imposed by the assumption that
null particles have uniform directions in the tangent plane at the
inner radius.

\section{Conclusion and Future Work}

The Orbiting Radiation star is a compact star based on null particles
capable of explaining large-mass gravitational wells of any scale.
The constituent null particles could be something exotic or something
more common, like photons.  The \OR\ is observationally distinct from
most other compact stars because test-particle photons entering the
\OR\ from infinity eventually emerge.

One element not addressed here is formation.  Could an \OR\ be created
by a natural or plausible designed process?  It is plausible that
variations of the \OR\ can consist of null particles with a variety of
impact parameters rather than a single impact parameter since existing
solutions can trap photons with other impact parameters as in
Figure~\ref{fig:trapped}.  As a consequence, a plausible formation
process may be somewhat ``dirty'', involving null particles with
imprecise impact parameters.  However, modeling the dynamics of
formation and general stability requires substantially more thought.

\section*{Appendix }
\label{sec:appendix}
\subsection*{proof of Lemma~\ref{lemma:SDT}}
First we prove the following:
\begin{lemma}\label{lemma:SD}  For null dust (i.e. assuming $R \equiv g_{\nu\nu}R^{\nu\nu} = 0$)
we can define $S \equiv A'/A + B'/B$ and $D\equiv B'/B - A'/A$ as the sum and difference
of $\log(B)'$ and $\log(A)'$, and get the following simplifications:
\begin{eqnarray}
\label{eq:sum}
S  & = &   r  R_{rr} + r B R_{tt}/A \\
\label{eq:diff}
D & = & (2 B R_{\theta\theta} - 2B + 2)/r\\
A' & = & A(S - D)/2 \\
B' & = & B(S + D)/2
\end{eqnarray}
\end{lemma}
\noindent
{\bf Proof:} Starting with:
\begin{eqnarray*}
R_{\theta\theta} &=& 1 - \frac{1}{B}  - \frac{r}{2B}\left(\frac{A'}{A}
- \frac{B'}{B}\right) \\
R_{\theta\theta} &=& 1 - \frac{1}{B}  + \frac{rD}{2B} \\
2B R_{\theta\theta} &=& 2B - 2  + rD \\
2(B R_{\theta\theta} +1 - B)/r &=& D
\end{eqnarray*}
To derive our equation $S$ we start with $R_{tt}$ and $R_{rr}$:
\begin{eqnarray*}
R_{tt} &=& \frac{A''}{2B} + \frac{A'}{rB} - \frac{A'}{4B}\left(\frac{A'}{A} + \frac{B'}{B}\right) \\
R_{rr} &=& -\frac{A''}{2A} + \frac{B'}{rB} + \frac{A'}{4A}\left(\frac{A'}{A} + \frac{B'}{B}\right)
\end{eqnarray*}
Substituting in $S$ and multiplying by $4B$ and $4A$ respectively we get:
\begin{eqnarray*}
4 B R_{tt} &=&  2A'' + 4 A'/r - A'S \\
4 A R_{rr} &=& -2A'' + 4 AB'/(rB) + A'S
\end{eqnarray*}
Adding these together we get:
\begin{eqnarray*}
4 (A R_{rr} +B R_{tt}) &=&  4 A'/r +  4 AB'/(rB) \\
 r (A R_{rr} + B R_{tt})/A &=&   A'/A +   B'/B \\
 r (A R_{rr} + B R_{tt})/A &=&  S
\end{eqnarray*}
\hfill $\Box$

\noindent
{\bf Proof of lemma \ref{lemma:SDT}:}
Define $S \equiv A'/A + B'/B$ and $D\equiv B'/B - A'/A$ as in lemma
\ref{lemma:SD}, and using what we learned there:
\begin{eqnarray*}
S  & = &   r  R_{rr} + r B R_{tt}/A \\
  & = &   rB(  R_{rr}/B +  R_{tt}/A) \\
  & = &   rB(  R^{rr}B +  R^{tt}A) \\
  & = &   8\pi G rB(  T^{rr}B +  T^{tt}A) \\
\end{eqnarray*}
Using the fact that $T$ is traceless, so $- AT^{tt} + BT^{rr} + 2 r^2
T^{\theta\theta} = 0$ we get:
\begin{eqnarray*}
S  & = &   8\pi G rB(  2 A T^{tt} - 2 r^2 T^{\theta\theta})
\end{eqnarray*}
Continuing with $D$:
\begin{eqnarray*}
D & = & (2 B R_{\theta\theta} - 2B + 2)/r\\
  & = & (2 B R^{\theta\theta}r^4 - 2B + 2)/r\\
  & = & (16 \pi G Br^4 T^{\theta\theta} - 2B + 2)/r\\
S+D& = & 16 \pi G rBA T^{tt} - 2B/r + 2/r\\
(S+D)/2& = & 8 \pi G rBA T^{tt} - B/r + 1/r\\
(S-D)/2& = & - 16 \pi G Br^3 T^{\theta\theta} + 8 \pi G rBA T^{tt} + B/r - 1/r
\end{eqnarray*}
\hfill $\Box$

\subsection*{Proof of Theorem \ref{thm:three}}

This section proves theorem \ref{thm:three} which derives the
 limiting solution to our equations when $\impact = 1$ and $\flux
 \to \flux_0$.  Define,
\begin{equation}
\label{eq:3}
F = \frac{r^2}{\flux} \sqrt{1 - A / r^2}
\end{equation}

\begin{lemma}\label{lemma:EB}
Our ODEs are:
\begin{eqnarray*}
3r^4  + \flux^2 F^2  - 2\flux^2 FF'r  & = &  B(\flux^2 F + r^4 -
 \flux^2 F^2) \\
r B' & = &   \frac{B^2}{(1 - \flux^2 F^2/r^4)F} - B^2 + B
\end{eqnarray*}
\end{lemma}

\noindent{\bf Proof of Lemma \ref{lemma:EB}:}
Taking $\impact = 1$ we have:
\begin{eqnarray*}
r A'/A & = &  \frac{\flux B \sqrt{1 -  A/r^2}}{A} +
B - 1 \\
r B'/B & = &   \frac{\flux B}{A\sqrt{1 - A/r^2}} - B + 1
\end{eqnarray*}
From our definition of $F \equiv \frac{r^2}{\flux} \sqrt{1 - A / r^2}$ we get:
\begin{eqnarray*}
 A & = & r^2 - \flux^2 F^2/r^2 \\
 A' & = & 2r - 2\flux^2 FF'r^{-2} + 2\flux^2 F^2r^{-3}\\
r A' & = &  \flux B \sqrt{1 -  A/r^2} + (B-1) A \\
2r^2 - 2\flux^2 FF'r^{-1} + 2\flux^2 F^2r^{-2}  & = &  \flux^2 BF/r^2 + (B-1)(r^2 - \flux^2 F^2/r^2) \\
2r^4 - 2\flux^2 FF'r + 2\flux^2 F^2  & = &  \flux^2 BF + (B-1)(r^4 - \flux^2 F^2) \\
2r^4 - 2\flux^2 FF'r + 2\flux^2 F^2  & = &  \flux^2 BF + Br^4 - B\flux^2 F^2-r^4 + \flux^2 F^2 \\
3r^4  + \flux^2 F^2  - 2\flux^2 FF'r  & = &  B(\flux^2 F + r^4 -
 \flux^2 F^2) 
\end{eqnarray*}
Now working on the second ODE:
\begin{eqnarray*}
r B'/B & = &   \frac{\flux B}{A\sqrt{1 - A/r^2}} - B + 1\\
r B'/B & = &   \frac{B\flux}{(r^2 - \flux^2 F^2/r^2)F\flux/r^2} - B + 1\\
r B' & = &   \frac{B^2}{(1 - \flux^2 F^2/r^4)F} - B^2 + B
\end{eqnarray*}
\hfill $\Box$

\begin{lemma}\label{lemma:poly:4} If $\lim_{r\to\infty} B = 3$, then
$F$ and $B$ can be written as polynomials in $r^{-4}$.
\end{lemma}

\noindent
{\bf Proof of Lemma \ref{lemma:poly:4}:}

Call the space spanned by polynomials in $r^{-4}$, ${\cal P}_4$.  Define
$x = r^{-4}$, then we are claiming that $F = F_0 + F_1 x + F_2 x^2 +
\cdots$ and $B = 3 + B_1 x + B_2 x^2 + \cdots$.  Technically, what
this means is that $F$ and $B$ are analytic in $x$ in the neighborhood of
$x=0$.  We won't take that route.  Instead, we show that writing 
\begin{displaymath}
F = F_0 + F_1 r^{-4} + F_2 r^{-8} + \cdots
\end{displaymath}
and
\begin{displaymath}
B = 3 + B_1 r^{-4} + B_2 r^{-8} + \cdots
\end{displaymath}
allows us to solve our ODE's without introducing terms of the form
 $r^{-4i +1}$, $r^{-4i +2}$ or $r^{-4i +3}$.

First notice that if $F \in {\cal P}_4$ then $r F' \in {\cal P}_4$.
Likewise, if $B \in {\cal P}_4$ then so is $r B' \in {\cal P}_4$.
We abuse notation and write any term which is in ${\cal P}_4$ as
${\cal P}_4$.  So, ${\cal P}_4+{\cal P}_4={\cal P}_4$ and ${\cal
P}_4*{\cal P}_4 = {\cal P}_4$.  So,
\begin{eqnarray*}
3r^4  + \flux^2 F^2  - 2\flux^2 FF'r  & = &  B(\flux^2 F + r^4 -
 \flux^2 F^2) \\
3r^4  + \flux^2 F^2  - 2\flux^2 FF'r  & = &  B\flux^2 F + Br^4 -
 B\flux^2 F^2 \\
3r^4  + {\cal P}_4\times{\cal P}_4  - {\cal P}_4\times{\cal P}_4  & = &  {\cal P}_4\times{\cal P}_4 + Br^4 -
{\cal P}_4\times{\cal P}_4\times{\cal P}_4 \\
{\cal P}_4   & = & (B-3) r^4
\end{eqnarray*}
But, since $B-3$ has a constant term which is $0$, we see that
$(B-3)r^4 \in {\cal P}_4$.  

If we now consider our second ODE,
\begin{eqnarray*}
r B' & = &   \frac{B^2}{(1 - \flux^2 F^2/r^4)F} - B^2 + B \\
r B'+B^2 - B & = &   \frac{B^2}{(1 - \flux^2 F^2/r^4)F}\\
F(1 - \flux^2 F^2/r^4)(r B'+B^2 - B) & = &  B^2\\
\end{eqnarray*}
All terms are in ${\cal P}_4$.

\hfill $\Box$

\begin{lemma}\label{lemma:equations} If $F$ and $B$ are in ${\cal
P}_4$ and $B_0=3$, then:
\begin{eqnarray*}
B_0 & = & 3 \\
B_1    & = &  \flux^2 (F_0^2 -  B_0(F_0 -
 F_0^2)) \\
B_2 &= & \flux^2 (2 F_0 F_1  + 8 F_0 F_1 +
2 B_0 F_0 F_1 + B_1 F_0^2 -  B_0 F_1 - B_1 F_0)\\
&\vdots& \\
F_0 & = & B_0/(B_0 - 1)\\
F_1 &=&\frac{2B_0 B_1 - F_0(\flux^2 F_0^2(B_0^2 - B_0) -4B_1 +2 B_0 B_1 - B_1) }{B_0^2 - B_0}\\
&\vdots& \\
\end{eqnarray*}
\end{lemma}

\noindent{\bf Proof of lemma \ref{lemma:equations}:}

\begin{eqnarray*}
F &=& F_0 + F_1 r^{-4} + F_2 r^{-8} + \cdots \\
rF'& =&0   -4 F_1 r^{-4} -8 F_2 r^{-8} + \cdots\\
B &=& 3 + B_1 r^{-4} + B_2 r^{-8} + \cdots \\
rB'& =&0   -4 B_1 r^{-4} -8 B_2 r^{-8} + \cdots
\end{eqnarray*}

\begin{eqnarray*}
3r^4  + \flux^2 F^2  - 2\flux^2 FF'r  & = &  B(\flux^2 F + r^4 -
 \flux^2 F^2) 
\end{eqnarray*}
\begin{displaymath}
Br^4 = 3r^4  + \flux^2 F^2  - 2\flux^2 FF'r +
B \flux^2 F^2 -  B\flux^2 F
\end{displaymath}
extracting the $r^4$ terms yields $B_0 = 3$.
Extracting the constant terms yields:
\begin{displaymath}
B_1     =   \flux^2 (F_0^2 -  B_0(F_0 -  F_0^2) )
\end{displaymath}
Continuing with the $r^{-4}$ terms:
\begin{displaymath}
B_2 =  \flux^2 (2 F_0 F_1  + 8 F_0 F_1 +
2 B_0 F_0 F_1 + B_1 F_0^2 -  B_0 F_1 - B_1 F_0)
\end{displaymath}

Proceeding in this fashion we get a series of equations involving
higher and higher order terms for $B_i$.  

Now using the other ODE for $B$ we get:
\begin{eqnarray*}
r B' & = &   \frac{B^2}{F (1-\flux^2 F^2/r^4)} - B^2 + B\\
r F B' & = &   \frac{B^2}{(1-\flux^2 F^2/r^4)} - FB^2 + FB\\
r F B'+FB^2 - FB & = &   \frac{B^2}{(1-\flux^2 F^2/r^4)}\\
(1-\flux^2 F^2/r^4)(r F B'+FB^2 - FB) & = &   B^2 \\
F(1-\flux^2 F^2/r^4)(r  B'+B^2 - B) & = &   B^2 
\end{eqnarray*}
Extracting the constant terms out of this we get:
\begin{eqnarray*}
(F_0)(1)(0 + B_0^2 - B_0) & = & B_0^2\\
F_0 B_0^2 - F_0 B_0 & = & B_0^2 \\
F_0 B_0 - F_0  & = & B_0 \\
F_0 & = & B_0/(B_0 - 1)
\end{eqnarray*}
Extracting the $r^{-4}$ terms (This means extracting two linear terms
from the product along with one $r^{-4}$, which can be done in three
ways.): 
\begin{eqnarray*}
(F_1)(1)(0 + B_0^2 - B_0) && \\
+ (F_0)(\flux^2 F_0^2)(0 + B_0^2 - B_0) && \\
+ (F_0)(1)(-4B_1 +2 B_0 B_1 - B_1) &=& 2B_0 B_1 
\end{eqnarray*}
\begin{displaymath}
F_1(B_0^2 - B_0) + F_0(\flux^2 F_0^2(B_0^2 - B_0) -4B_1 +2 B_0 B_1 - B_1) = 2B_0 B_1 
\end{displaymath}
\begin{displaymath}
F_1(B_0^2 - B_0) =2B_0 B_1 - F_0(\flux^2 F_0^2(B_0^2 - B_0) -4B_1 +2 B_0 B_1 - B_1) 
\end{displaymath}
\begin{displaymath}
F_1 =\frac{2B_0 B_1 - F_0(\flux^2 F_0^2(B_0^2 - B_0) -4B_1 +2 B_0 B_1 - B_1) }{B_0^2 - B_0}
\end{displaymath}

\hfill $\Box$

\begin{lemma}\label{lemma:coef} Solving the equations in
lemma \ref{lemma:equations} we get:
\begin{eqnarray*}
B_0 & = & 3 \\
F_0 & = & 3/2 \\
B_1    & = &  \flux^2 9/2 \\
F_1 & = &  0 \\
B_2 & = &  \flux^4 27/8.
\end{eqnarray*}
\end{lemma}

\noindent{\bf Proof:}
We have $B_0  =  3$, so substituting this in to $F_0$:
\begin{displaymath}
F_0  =  3/(3 - 1) = 3/2
\end{displaymath}
Plugging these both in to $B_1$:
\begin{displaymath}
B_1     =   \flux^2 ((3/2)^2 -  3((3/2) - (3/2)^2)) = \flux^2 9/2
\end{displaymath}
Plugging all three into $F_1$:
\begin{eqnarray*}
F_1 &=&\frac{6 (\flux^2 9/2) - (3/2)(\flux^2 (3/2)^2 6 -4(\flux^2 9/2) +6 (\flux^2 9/2) - (\flux^2 9/2)) }{6}\\
F_1 &=& \flux^2\frac{27 - (3/2)(27/2 - 18 + 27 - 9/2) }{6}\\
F_1 &=& 0
\end{eqnarray*}
Likewise:
\begin{eqnarray*}
B_2 &= & \flux^2 ((\flux^2 9/2) (3/2)^2 - (\flux^2 9/2) (3/2))\\
B_2 &= & \flux^4 27/8
\end{eqnarray*}

\hfill$\Box$

\noindent{\bf Proof of Theorem \ref{thm:three}:}

We have 
\begin{eqnarray*}
B & = & 3 + (9/2) \flux^2/r^4 + (27/8) \flux^4/r^8 + O(r^{-12}) \\
F & = & 3/2 +O(r^{-8}) \\
A & = & r^2 - \flux^2 F^2/r^2 \\
A & = & r^2 - (9/4)\flux^2/r^2 + O(r^{-10})
\end{eqnarray*}

\hfill$\Box$

\subsection*{Proof of Theorem \ref{thm:nine}}
Before proving this theorem, we first introduce a change of
variables:
\begin{eqnarray}
z & = & \flux^2(r-1) \label{eqn:z}\\
H & = &  \sqrt{2z - \flux^2(A - 1)} \label{eqn:H}
\end{eqnarray}
We are assuming that the inner turning point is set equal to $1$ and
that $A(1) = 1$.  The following lemma (whose proof is given later since it
is just algebra) converts our ODEs to these variables under the
assumption that $\flux \approx \infty$.
\begin{lemma}\label{lemma:zH} Define $z$ and $H$ as in (\ref{eqn:z})
and (\ref{eqn:H}).  Then after taking the limit as $\flux \to
 \infty$ we get:
\begin{eqnarray}
  dH/dz & = &(3/H - B  - B/H)/2 \label{eqn:dH} \\
  dB/dz & = & B^2/H \label{eqn:dB}
\end{eqnarray}
\end{lemma}

Proof of Lemma \ref{lemma:zH}:

Our original ODEs are (with $\impact=1$):
\begin{eqnarray*}
r A'/A & = &  \frac{\flux B \sqrt{1 -  A/r^2}}{A} +
B - 1 \\
r B'/B & = &   \frac{\flux B}{A\sqrt{1 - A/r^2}} - B + 1
\end{eqnarray*}
Proof of the change of variables:
\begin{eqnarray*}
dz & = & \flux^2 dr\\
dB/dz & = & B'/\flux^2  \\
% dH/dz & = & (1/2)(2 - A')/\sqrt{2z - \flux^2(A - 1)} \\
dH/dz & = & (1 - A'/2)/H
\end{eqnarray*}
This leads to
\begin{eqnarray*}
A' & = & 2 - 2HdH/dz \\
A & = & 1+(2z - H^2)/\flux^2\\
r & = & 1 + z/\flux^2 \\
\flux r\sqrt{1 - A/r^2} & = &  \flux \sqrt{r^2 - A} \\
 & = &  \flux \sqrt{(1+z/\flux^2)^2 - 1 - (2z - H^2)/\flux^2} \\
 & = &  \flux \sqrt{ z^2/\flux^4 + H^2/\flux^2} \\
 & = &  \sqrt{H^2 +  z^2/\flux^2} \\
\end{eqnarray*}
Working these into our first ODE:
\begin{eqnarray*}
A' & = &  \frac{\flux B \sqrt{1 -  \impact^2A/r^2}}{r} +
BA/r - A/r \\
2 - 2HdH/dz & = & (B/r^2) \sqrt{H^2 + z^2/\flux^2} + (B-1)A/r\\
2 - 2HdH/dz & = & (B/r^2) \sqrt{H^2 + z^2/\flux^2} + (B-1)\frac{(1+(2z
- H^2)/\flux^2)}{1 + z/\flux^2}
\end{eqnarray*}
Now taking the limit as $\flux \to \infty$ 
\begin{eqnarray*}
2 - 2HdH/dz & = & B \sqrt{H^2} + B - 1\\
 - 2HdH/dz & = & B H + B - 3\\
  dH/dz & = &3/(2H) - B/2  - B/(2H)
\end{eqnarray*}
Now trying the second ODE:
\begin{eqnarray*}
dB/dz & = &   \frac{B^2}{A\flux r \sqrt{1 - \impact^2A/r^2}} - B^2/r\flux^2 + B/r\flux^2\\
dB/dz & = &   \frac{B^2}{A \sqrt{H^2 + z^2/\flux^2}} - B^2/r\flux^2 + B/r\flux^2\\
dB/dz & = &   \frac{B^2}{(1+(2z - H^2)/\flux^2) \sqrt{H^2 + z^2/\flux^2}} - B^2/((1 + z/\flux^2)\flux^2) + B/(( 1 + z/\flux^2)\flux^2)
\end{eqnarray*}
Now taking the limit as $\flux \to \infty$ 
\begin{eqnarray*}
dB/dz & = &   \frac{B^2}{H}\\
\end{eqnarray*}
So our equations are:
\begin{eqnarray*}
2 H dH/dz & = & 3 - BH - B \\
H dB/dz & = & B^2
\end{eqnarray*}

\hfill$\Box$

\vspace{2em}
\noindent
{\bf Proof of Theorem \ref{thm:nine}:}
Taking the ratio of (\ref{eqn:dH}) to (\ref{eqn:dB}) we get:
\begin{displaymath}
 \frac{dH}{dB} = (3/B^2 - H/B - 1/B)/2
\end{displaymath}
Which can be solved as for some constant k.\footnote{a check:
\begin{eqnarray*}
-k B^{-3/2} + 6/B^2 &=& 3/B^2 -k B^{-3/2} + 1/B + 3/B^2 - 1/B
\end{eqnarray*}}
\begin{eqnarray*}
H &= &k/\sqrt{B} -1 - 3/B \\
& = & -3(1/\sqrt{B} - a/3)(1/\sqrt{B} - 1/a)
\end{eqnarray*}
where $k = a + 3/a$.  

In the original variables the turning point condition is $b^2 A =
r^2$.  This corresponds to a turning point condition is
$H=0$.\footnote{If $H = 0$, then $2z = \flux^2(A-1)$ which is the same
as $2(r-1) = A - 1$.  Which is $A =2r -1 = r^2 - z^2/\flux^4$.  So as
$\flux \to \infty$ we see that $H=0$ are the turning points.}
  The turning points are then:
\begin{eqnarray*}
B_{\hbox{inner}} &=& a^2\\
B_{\hbox{outer}}& =& 9/a^2 = 9/B_{\hbox{inner}}
\end{eqnarray*}

For the flat interior case, we have one root at $B=1$ and hence the
 second root is at $B=9$ which means $k=4$:
\begin{displaymath}
H = 4/\sqrt{B} - 1 - 3/B
\end{displaymath}
Since we know the geometry is Schwarzschild outside $r_{\text{outer}}$,
and $B=9 = 1/(1-2/r_{\text{outer}})$, we can deduce that
$r_{\text{outer}} = 9/8$.  By the construction we see that
$r_{\text{inner}} \simeq r_{\text{outer}}$.  This completes our claims.
\hfill$\Box$

Notice that if we allow negative mass solutions on the inside, any
combination works out for which $B_i B_o = 9$.  Alternatively, if we
consider a black hole at the center, then we can have a heavy shell of
``orbiting radiation'' at any distance from 2.25 out to 3.0.  This
shell provides the extra mass necessary to keep them in orbit.


\begin{thebibliography}{1}

\bibitem{crossstreaming} J Bicak and P Hajicek, ``Canonical theory of spherically symmetric spacetimes with cross-streaming null dusts'', Physical Review D, 2003.

\bibitem{BS86} M Colpi, SL Shapiro, and I Wasserman, ``Boson stars:
  Gravitational equilibria of self-interacting scalar fields'',
  Physical review letters, 1986.

\bibitem{code}D Foster and J Langford, OR-star code, \url{https://github.com/JohnLangford/OR-star}.

\bibitem{Gergeley} LA Gergely, ``Spherically symmetric static solution for colliding null dust'', \url{http://arxiv.org/abs/gr-qc/9809024}, 1998.

\bibitem{PhotonStars} H-J Schmidt and F Homann, ``Photon Stars'',
  General Relativity and Gravitation 32, page 919, 2000.

\bibitem{BS92}P Jetzer, ``Boson stars'', Physics Reports, 1992.

\bibitem{KC2010} J Kijowski and E Czuchry, ``Dynamics of a self gravitating light-like shell with spherical symmetry'', Classical and Quantum Gravity, 2010.

\bibitem{Grava2001} PO Mazur and E Mottola, ``Gravitational condensate stars: an alternative to black holes'', \url{http://arxiv.org/pdf/gr-qc/0310107.pdf}.

\bibitem{Anisotropic2007} R Sharma and SD Maharaj, MNRAS 375, 1265-1268, 2007.

\bibitem{Tolman} R Tolman, ``Relativity, Thermodynamics, and Cosmology'', Oxford University Press, Clarendon Press, 1934.

\bibitem{Vaidya} PC Vaidya, ``Radiating Spherical Star'', Nature (London) 171, page 260, 1953.

\bibitem{Grava2004} M Viser and DL Wiltshire, ``Stable Gravastars-an alternative to black holes?'', Classical and Quantum Gravity, 2004.

\bibitem{Geons} JA Wheeler, ``Geons'', Phys. Rev. 97, 511, January
  1955.

\bibitem{Zee} A Zee (2013). Einstein gravity in a nutshell. Princeton University Press.


\end{thebibliography}
\end{document}